\documentclass[conference]{IEEEtran}
\IEEEoverridecommandlockouts
\usepackage{cite}
\usepackage{amsmath,amssymb,amsfonts}
\usepackage{graphicx}
\usepackage{textcomp}
\usepackage{xcolor}
\usepackage{url}
\usepackage{algorithm}
\usepackage{algpseudocode}
\usepackage{subfigure} 
\usepackage{pifont}
\usepackage{booktabs, multirow}
\newcommand{\var}{\texttt}

\usepackage{diagbox}    
\usepackage{colortbl}
\newcommand{\cmark}{\text{\ding{51}}}
\newcommand{\xmark}{\text{\ding{55}}}

\usepackage{makecell}

\def\BibTeX{{\rm B\kern-.05em{\sc i\kern-.025em b}\kern-.08em
    T\kern-.1667em\lower.7ex\hbox{E}\kern-.125emX}}

\begin{document}

\title{BABD: A Bitcoin Address Behavior Dataset for Pattern Analysis\\

\thanks{The first three authors contributed equally to this work.}
\thanks{Corresponding author at: School of Computer Science, China University of Geosciences, Wuhan, China. Email: weirencs@cug.edu.cn.}
}
\author{\IEEEauthorblockN{
Yuexin Xiang\IEEEauthorrefmark{1},
Yuchen Lei\IEEEauthorrefmark{1}, 
Ding Bao\IEEEauthorrefmark{1},
Wei Ren\IEEEauthorrefmark{1}\IEEEauthorrefmark{2}\IEEEauthorrefmark{3},
Tiantian Li\IEEEauthorrefmark{1}, 
Qingqing Yang\IEEEauthorrefmark{1},\\ 
Wenmao Liu\IEEEauthorrefmark{4},
Tianqing Zhu\IEEEauthorrefmark{5},
and
Kim-Kwang Raymond Choo\IEEEauthorrefmark{6}}
\IEEEauthorblockA{\IEEEauthorrefmark{1}School of Computer Science, China University of Geosciences, Wuhan 430074, China}
\IEEEauthorblockA{\IEEEauthorrefmark{2}Henan Key Laboratory of Network Cryptography Technology, Zhengzhou 450001, China}
\IEEEauthorblockA{\IEEEauthorrefmark{3}Hubei Key Laboratory of Intelligent Geo-Information Processing, Wuhan 430074, China}
\IEEEauthorblockA{\IEEEauthorrefmark{4}NSFOCUS Technologies Group Co.,Ltd.}
\IEEEauthorblockA{\IEEEauthorrefmark{5}School of Computer Science, University of Technology Sydney, Sydney, Australia}
\IEEEauthorblockA{\IEEEauthorrefmark{6}Department of Information Systems and Cyber Security, University of Texas at San Antonio \\ San Antonio, TX 78249-0631, USA}}

\maketitle

\begin{abstract}
Cryptocurrencies are no longer just the preferred option for cybercriminal activities on darknets, due to the increasing adoption in mainstream applications. This is partly due to the transparency associated with the underpinning ledgers, where any individual can access the record of a transaction record on the public ledger. In this paper, we build a dataset comprising Bitcoin transactions between 12 July 2019 and 26 May 2021. This dataset (hereafter referred to as BABD-13) contains 13 types of Bitcoin addresses, 5 categories of indicators with 148 features, and 544,462 labeled data, which is the largest labeled Bitcoin address behavior dataset publicly available to our knowledge. We then use our proposed dataset on common machine learning models, namely: \emph{k}-nearest neighbors algorithm, decision tree, random forest, multilayer perceptron, and XGBoost. The results show that the accuracy rates of these machine learning models for the multi-classification task on our proposed dataset are between 93.24\% and 97.13\%. We also analyze the proposed features and their relationships from the experiments, and propose a \emph{k}-hop subgraph generation algorithm to extract a \emph{k}-hop subgraph from the entire Bitcoin transaction graph constructed by the directed heterogeneous multigraph starting from a specific Bitcoin address node (e.g., a known transaction associated with a criminal investigation). Besides, we initially analyze the behavior patterns of different types of Bitcoin addresses according to the extracted features.
\end{abstract}


\section{Introduction}
Cryptocurrencies, such as Bitcoin, remain increasingly popular. For example, according to CoinMarketCap, the global cryptocurrency market capital is \$1.77 Trillion U.S. dollars and Bitcoin's market capitalization is estimated to be at \$0.74 Trillion U.S. dollars (as of February 27, 2022 - see \url{https://coinmarketcap.com}). Similar to fiat currencies, there are concerns about the misuse of cryptocurrencies (e.g., cybercriminal and darknet markets). However, unlike fiat currencies, one can more easily trace cryptocurrency transactions partly because of the transparency nature of the public ledger.

It is, therefore, unsurprising that there have been numerous attempts to design techniques to facilitate cryptocurrency transaction tracing \cite{Alqassem,Nerurkar,Popuri,Ron,Serena,Tao,tao2021complex}. For example, in the context of Bitcoin transactions, one could utilize graph analysis to determine or classify different address types. There are, however, a number of challenges and limitations in existing approaches, ranging from accuracy to information loss to (in)completeness (e.g., not considering internal relations and differences of different classifications). Another limitation we observe is the lack of a comprehensive dataset that can be used as a baseline for the research community. 

Therefore, in this paper, we present a general framework that can be used to build a Bitcoin transaction graph by the directed heterogeneous multigraph and analyze address nodes from the graph. We then collect and compile Bitcoin transactions that occur between 12 July 2019 and 26 May 2021 into a dataset (hereafter referred to as BABD-13), which can be found on Kaggle\footnote{ \url{https://www.kaggle.com/datasets/lemonx/babd13}}. The latter has 13 types of Bitcoin addresses, 544,462 labeled data, and 5 categories of indicators with 148 features. We then use BABD-13 as the baseline dataset to evaluate common machine learning models (i.e., \emph{k}-nearest neighbors algorithm, decision tree, random forest, multilayer perceptron, and XGBoost). Findings from the evaluations suggest that the accuracy obtained ranges between 93.24\% and 97.13\%.
To extract the structural features, we also propose a novel method to generate \emph{k}-hop subgraph for an address node from the entire Bitcoin transaction graph built by the directed heterogeneous multigraph. Finally, we simply summarize the several behavior patterns of different Bitcoin addresses according to the experimental results as the basis of future work.

In the next section, we will briefly review the extant literature.

\section{Related Work}
\label{sec:relatedwork}

As discussed earlier, a number of approaches to analyze Bitcoin transactions have been proposed in the literature, such as those focusing on quantitative studies \cite{Ron}. For example, using Bitcoin transaction data from 2009 to 2014, Alqassem et al. \cite{Alqassem} studied the evolution of graph structural properties over time. They observed that the Bitcoin transaction graph is generally similar to typical social networks in the structural index. To understand the structural features of the Bitcoin transaction graph, Popuri and Gunes \cite{Popuri} measured the general characteristics of the Bitcoin network using the complex network theory. Similarly, Serena et al. \cite{Serena} considered the cryptocurrency transaction graph as a complex network and studied the Bitcoin transaction graph using complex network-based methods. Specifically, using degree distribution and aggregation clustering coefficient, they calculated several simple indexes and their changes \cite{Serena}. More recently in 2021, Tao et al. \cite{Tao} proposed and implemented a complex network-based framework to comprehensively analyze the Bitcoin transaction graph. They observed that the non-rich club effect, small-world phenomenon, and other typical characteristics exist in the Bitcoin transaction graph.

Due to the increasing popularity of cryptocurrencies, there has been a corresponding increase in such currencies from governments and regulatory agencies. This reinforces the importance of understanding the full Bitcoin transaction graph so that governments and regulatory agencies (e.g., taxation, law enforcement, and financial intelligence units, as well as anti-money laundering/counter-terrorism financing regulators) can monitor and trace criminal proceeds. In other words, how do we more accurately distinguish normal from suspicious cryptocurrency addresses? Such addresses generally include exchange-related addresses and mining pool-related addresses. 

Ranshous et al. \cite{Ranshous} used the directed hypergraph to analyze the patterns associated with exchange addresses, and proposed different types of short thick bands (STBs) to identify the patterns. In addition, they also applied basic machine learning methods to classify early exchange addresses. Focusing on mining pools (another kind of normal pattern), Romiti et al. \cite{Romiti} selected three of four of the biggest pools to empirically investigate the relationships among them. They mainly utilized economic activity pattern-based ways to explore the relationships between miner-owned addresses and pools. In separate work, Tovanich et al. \cite{Tovanich} studied pool hopping behaviors in 15 pools of Bitcoin transactions. Based on the empirical study and their proposed heuristic algorithm designed to describe the payout flows, they determined those pool fees and payout schemes are the two most important factors to influence the behaviors of miner-owned addresses.

There have also been attempts to detect activities associated with illicit addresses or transactions \cite{Li,Paquet-Clouston1,Weber,Wu3}. For example, Liao et al. \cite{Liao} measured data such as Bitcoin amount in ransomware cases and the resulting financial loss. Conti et al. \cite{Conti} also proposed an approach to analyzing the financial impact of ransomware and presented the timeline of the ransomware process. They then built a public dataset of ransomware-related Bitcoin addresses. Additionally, Paquet-Clouston et al. \cite{Paquet-Clouston} proposed an efficient framework to identify and collect ransomware attack-related Bitcoin transaction addresses. Their research demonstrated the change of the Bitcoin amount over time from 2013 to 2017. Bartoletti et al. \cite{Bartoletti} collected Ponzi scheme Bitcoin addresses from various Bitcoin forums, and based on their analyses they proposed 11 features for Bitcoin address classification. Vasek and Moore \cite{Vasek} obtained their Ponzi scheme data from a bitcoin forum\footnote{\url{https://bitcointalk.org}} and observed that cybercriminals interacted more frequently with the victims and posted more comments in the thread. In addition, Toyoda et al. \cite{Toyoda} presented a systematic approach to study the high yield investment program (HYIP), which is a kind of Ponzi scheme. They analyzed the HYIP addresses based on historic records and applied supervised learning to classify the HYIP addresses. The accuracy rate of their approach is as high as 93.75\%. 

There have also been approaches designed to learn the similarities and differences among different kinds of Bitcoin addresses or transactions in the Bitcoin transaction graph \cite{Li,Liu1,Monaco,Chang,Zola,Nerurkar}. However, most existing approaches prefer to utilize the simplified Bitcoin transaction graph (e.g., undirected simple graph), rather than its original structure, in their analysis  \cite{Paquet-Clouston1}. Consequently, this may result in information loss. 

The types of different addresses in most existing works are inaccurate and/or incomplete. For instance, at most 7 types of addresses were analyzed together in recent works such as that of \cite{8751410}. This is clearly not sufficient in having an in-depth understanding of the behavior patterns in the Bitcoin transaction graph. 

In addition, the indicators utilized for analysis proposed in existing approaches, such as those of \cite{Monaco,Chang,Zola,Nerurkar}, are generally not systematic and/or comprehensive. Specifically, they do not classify their indicators and do not include key indicators that can be extracted from the Bitcoin transaction graph. Besides, these approaches are hard to replicate since it is not clear (publicly) how the Bitcoin transaction graph(s) is/are built and generated. 

In the next section, we will briefly introduce the relevant background materials.

\section{Background Materials}

\subsection{Bitcoin Graph Structure}
Bitcoin transaction graph is essentially a kind of unspent transaction output (UTXO) model. Aiming at different kinds of issues existing in the huge Bitcoin network, various Bitcoin network models are proposed by researchers \cite{Liu1,Wu2}. For a general analysis of the Bitcoin network, transaction nodes or address nodes are the main objects \cite{Bartoletti,Li,Monaco,Zheng2}, especially transaction nodes because they include more information. Besides, there are also a great number of works that regard related addresses as an entity for analysis \cite{Chang,Yousaf}.

\begin{figure}[!htbp]
  \centering
  \includegraphics[width=1\linewidth]{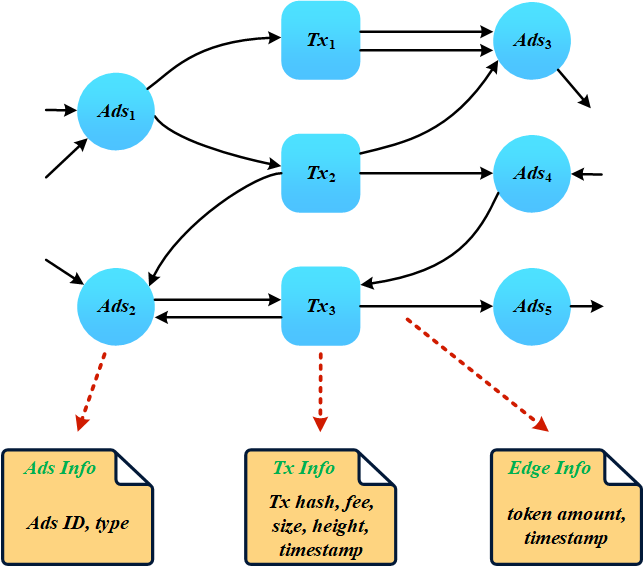}
  \caption{Bitcoin transaction graph structure}
  \label{BTC_structure}
\end{figure}

However, in our view, the above and similar approaches to simplify the Bitcoin transaction graph will lead to information loss in the Bitcoin graph analysis. In this paper, for diminishing the information loss of the Bitcoin network as much as possible while analyzing the transaction patterns, we proposed an improved directed heterogeneous multigraph Bitcoin structure based on the structure proposed by Maesa et al. \cite{Maesa}, which includes characteristics of both the address (\emph{Ads}) node and the transaction (\emph{Tx}) node to ensure the accuracy of the analysis. Our designed structure is shown in Fig. \ref{BTC_structure} in detail.

\subsection{Behavior Classification and Definition}
It is significant to classify and define typical behaviors of Bitcoin addresses exactly. Specifically, the concepts and basic motivations of different illicit and licit Bitcoin addresses will help us improve the effect of classification tasks for Bitcoin addresses and understand deeper the patterns that exist in Bitcoin address behaviors. 

We will study and analyze the mainstream 13 types of illicit and licit Bitcoin addresses in this paper, which are listed as follows:

\begin{enumerate}
    \item \textbf{\emph{Blackmail.}} Cryptocurrency blackmail has three typical categories that are ransomware, sextortion, and scam. Apart from the blackmail methods, these blackmail types are similar in most aspects, which utilize some ways to threaten or deceive the victims to pay a certain amount of cryptocurrency to several specific addresses.
    
    \item \textbf{\emph{Cyber-Security Service.}} Cyber-security services denote that the providers can offer payment gateways, proxy or virtual private network (VPN) services, and other cyber-security-related services. In this case, providers only accept cryptocurrency as payment in order to improve the security of their services. 
    
    \item \textbf{\emph{Darknet Market.}} Darknet markets are the markets hiding in the darknet, where people are able to buy and sell illegal stuff and services such as automatic rifles and assassination services anonymously. Besides, in order to enhance the anonymity of these dirty transactions, traders are oriented to exchange through cryptocurrencies.
    
    \item \textbf{\emph{Centralized Exchange.}}  Centralized cryptocurrency exchanges earn fees by acting as trustworthy intermediaries among their customers. They execute the ``Know Your Customer (KYC)" policy and allow their customers to trade cryptocurrency/cryptocurrency pairs (e.g., BTC/USDT and ETH/BTC pairs) and do swaps between cryptocurrency and fiat.
    
    \item \textbf{\emph{P2P Financial Infrastructure Service.}} P2P financial infrastructure services are the P2P financial activities that conduct only by cryptocurrency. Examples are bond markets and P2P lending platforms.
    
    \item \textbf{\emph{P2P Financial Service.}} P2P financial applications use cryptocurrency as a reward according to users' contributions to encourage users to complete more missions that are helpful to the organizers such as increasing the number of clicks of a concrete advertisement. This kind of service includes faucets, video sharing, and affiliate marketers.
    
    \item \textbf{\emph{Gambling.}} Cryptocurrency gambling denotes playing casino games, such as blackjack and roulette, in which only cryptocurrencies can be utilized as wagers.

    \item \textbf{\emph{Government Criminal Blacklist.}} Government criminal blacklist includes cryptocurrency addresses highly suspected or confirmed involved in criminal activities according to different countries' laws.

    \item \textbf{\emph{Money Laundering.}} Money Laundering is trading the dirty cryptocurrency in dirty addresses obtained from illegal transactions or activities to normal addresses in several complicated ways that others are hard to trace. Generally, money laundering includes the following three stages that are placement, laundering, and integration.

    \item \textbf{\emph{Ponzi Scheme.}} Ponzi schemes obey the rules that the initiator of the Ponzi scheme will pay high interests by cryptocurrency to the former investors through the investments from current involving investors. Therefore, the investors will believe that the initiator can easily earn profits for them in a related short time. 

    \item \textbf{\emph{Mining Pool.}} Cryptocurrency mining pools consist of a large number of miners who contribute their computational resources together for improving the possibility of finding a new block to gain the reward of cryptocurrencies. Thus, the majority of addresses related to coinbase transactions in Bitcoin belong to mining pools.

    \item \textbf{\emph{Tumbler.}} Cryptocurrency tumbler is a kind of service that prevents monitors or censors from tracing cryptocurrency flows according to the transparent ledger, which increases the anonymity of cryptocurrencies. Generally, the service providers combine multiple inputs during a long period of time and send different inputs to their planned destinations at random times to implement cryptocurrency tumblers. 
    \item \textbf{\emph{Individual Wallet.}} Individual wallets stand for the wallets owned by ordinary people who use cryptocurrency as a payment way in daily life, such as shopping and dining.
\end{enumerate}

\subsection {Data Collection}
There are two kinds of data we need to collect respectively that are Bitcoin ledger data and Bitcoin address data with labels. We will illustrate how we gather them in this section. 

\textbf{\emph{Bitcoin Ledger.}} We collect the Bitcoin ledger using the public API\footnote{\url{https://btc.com}}. Our current research on Bitcoin ledger data is based on 100,001 blocks data, where the block height is from 585,000 to 685,000 (from July 12, 2019 to May 26, 2021), including 516,167,131 address and transaction nodes and 713,703,239 relationship edges. Due to our device performance limitations, we only use 100,001 blocks to build the bitcoin transaction graph, while our approach can actually be able to build bitcoin transaction graphs of any size.

\textbf{\emph{Bitcoin Address.}} Compared to the Bitcoin ledger, it is harder to gather Bitcoin addresses with labels. In order to collect more data while keeping the accuracy of the data, we divide our collected data into two types are ``strong address (\emph{SA})" and ``weak address (\emph{WA})", where \emph{SA} denotes the address with the strongly confirmed label, in contract, \emph{WA} means the address with the week confirmed label (e.g. the reported address). \emph{SA}s mainly collected from the public dataset\footnote{\url{https://www.walletexplorer.com}} and government blacklists\footnote{\url{https://home.treasury.gov}}.  Furthermore, we also find valuable \emph{SA} data from the high influential papers \cite{Conti,Bartoletti}. Besides, \emph{WA}s are gathered from the public reported Bitcoin address dataset\footnote{\url{https://www.bitcoinabuse.com}} and a number of social accounts on Twitter\footnote{\url{https://twitter.com}}. All the data collected is listed in Table \ref{data}.

\begin{table}[!htbp]
\renewcommand\arraystretch{1.2}
\caption{Labeled bitcoin address} 
\label{data} 
\begin{tabular}{l l}
\noalign{\hrule height 0.5pt}
Type & Number (SA / WA) \\ 
\noalign{\hrule height 0.5pt} 
Blackmail & 8,686 (8 / 8,678)\\
Cyber-Security Service & 91,617 (91,617 / 0)\\
Darknet Market & 13,861 (13,861 / 0)\\
Centralized Exchange & 300,000 (300,000 / 0) \\
P2P Financial Infrastructure Service & 180 (180 / 0)\\
P2P Financial Service & 9,309 (9,309 / 0) \\
Gambling & 105,257 (105,257 / 0)\\
Government Criminal Blacklist & 27 (27 / 0)\\
Money Laundering & 16 (16 / 0)\\
Ponzi Scheme & 15 (8 / 7)\\
Mining Pool & 1,580 (1,580 / 0)\\
Tumbler & 12,412 (10,817 / 1,595)\\
Individual Wallet & 1,502 (52 / 1,450)\\
\noalign{\hrule height 0.5pt}
\textbf{\emph{Total Number}} & \textbf{544,462} (\textbf{532,732} / \textbf{11,730})\\
\noalign{\hrule height 0.5pt}
\end{tabular}
\end{table}

\section{Proposed Scheme}

\subsection{Framework}
Referring to the indicators and classification methods presented in several high influential related works \cite{Chang,Liu1,Monaco,Zola,Nerurkar,Liu,Greaves}, we proposed our own framework for extracting and analyzing Bitcoin address behaviors from the Bitcoin transaction graph that constructed using the directed heterogeneous multigraph. The proposed framework consists of two parts that are the \emph{\textbf{statistical indicator (SI)}} and the \emph{\textbf{local structural indicator (LSI)}} respectively.

\subsection{Statistical Indicator}
\emph{SI} in our proposed scheme is divided into four concrete indicator types, which we will briefly introduce as follows:
\begin{itemize}
    \item \textbf{\emph{Pure Amount Indicator (PAI).}} \emph{PAI} is related to the token amount attributes of \emph{Ads} nodes in the Bitcoin transaction graph. 
    \item \textbf{\emph{Pure Degree Indicator (PDI).}} \emph{PDI} includes the degree-related attributes of \emph{Ads} nodes in the Bitcoin transaction graph.
    \item \textbf{\emph{Pure Time Indicator (PTI).}} \emph{PTI} is time-related attributes of \emph{Ads} nodes.
    \item \textbf{\emph{Combination Indicator (CI).}} \emph{CI} is the combinations of the features from \emph{PAI}, \emph{PDI}, and \emph{PTI}.
\end{itemize}

The basic \emph{SI} is shown in Table \ref{NotationSI} that includes \emph{PAI}, \emph{PDI}, and \emph{PTI}, and features from \emph{CI} are illustrated in Table \ref{NotationCI}. In these tables, there are three points that we need to further explain. First, the token amount in the above tables is the original data of the Bitcoin transactions, i.e., BTC. Second, the basic unit mentioned in the above tables is the solar day of 24 hours. Finally, for the features of \emph{Ads} node in the above tables, we also compute their simplified features. That means we re-compute the same features after combining the same directed edges in the Bitcoin transaction graph with their attributes and make them newly added features (i.e., using PAIa11-R1 instead of PAIa11-1).

\begin{table*}[!htbp]
\centering
\renewcommand\arraystretch{1.5}
\caption{Basic statistical indicator} 
\label{NotationSI}
\begin{tabular}{l l l l}
\noalign{\hrule height 0.5pt}
Category & Notation & Description & Dataset\\
\noalign{\hrule height 0.5pt}
\textbf{\emph{PAI}} & $A_{a}^{in/out}$ & The input/output token amount of an address node & PAIa1\\ 
& $A_{a_{all}}^{in/out}$ & The total input/output token amount of an address node & PAIa11\\
& $DAa_1$ & The difference of $A_{a_{all}}^{in} - A_{a_{all}}^{out}$ & PAIa12\\
& $RAa_1$ & The ratio of $\frac{A_{a_{all}}^{in}}{A_{a_{all}}^{out}}$ & PAIa13\\
& $A_{a_{min/max}}^{in/out}$ & The minimum/maximum input/output token amount of an address node & PAIa14\\
& $DAa_2^{in/out}$ & The difference of $A_{a_{max}}^{in} - A_{a_{min}}^{in}$ and $A_{a_{max}}^{out} - A_{a_{min}}^{out}$ & PAIa15\\
& $RAa_2^{in/out}$ & The ratio of $\frac{DAa_2^{in}}{A_{a_{all}}^{in}}$ and $\frac{DAa_2^{out}}{A_{a_{all}}^{out}}$ & PAIa16\\
& $A_{a_{std}}^{in/out/all}$ & The standard deviation of all input/output/input and output token amounts of an address node & PAIa17\\
& $RAa_3^{in/out}$ & The ratio of $\frac{A_{a}^{in}}{A_{a_{all}}^{in}}$ and $\frac{A_{a}^{out}}{A_{a_{all}}^{out}}$ & PAIa2\\
& $RAa_{3_{min/max}}^{in/out}$ & The minimum/maximum input/output value of $RAa_3^{in/out}$ & PAIa21\\
& $RAa_{3_{std}}^{in/out}$ & The standard deviation of all input/output values of $RAa_3^{in/out}$ & PAIa22\\
\textbf{\emph{PDI}} & $A_d^{in/out/all}$ & The in-degree/out-degree/in-degree and out-degree of an address & PDIa1\\
& $RAd_1^{in/out}$ & The ratio of $\frac{A_d^{in}}{A_d^{all}}$ and $\frac{A_d^{out}}{A_d^{all}}$ & PDIa11\\
& $RAd_2$ & The ratio of $\frac{A_d^{in}}{A_d^{out}}$ & PDIa12\\
& $DAd_1$ & The difference of $A_d^{in} - A_d^{out}$ & PDIa13\\
\textbf{\emph{PTI}} & $A_p^{l}$ & The life period of an address & PTIa1\\
& $A_p^{v}$ & The active period of an address & PTIa2\\
& $RAp_1$ & The ratio of $\frac{A_p^v}{A_p^l}$ & PTIa21\\
& $A_n^v$ & The number of active times of an address in each basic unit of the active period & PTIa3\\
& $A_{n_{min/max/avg}}^v$ & The minimum/maximum/average active times of $A_n^v$ & PTIa31\\
& $DA_1^v$ & The difference of $A_{n_{max}}^v - A_{n_{min}}^v$ & PTIa32\\
& $A_{n_{std}}^v$ & The standard deviation of $A_n^{v}$ & PTIa33\\
& $A^{ti}$ & The time interval of transactions of an address (chronological order) & PTIa4\\
& $A_{min/max/avg}^{ti}$ & The minimum/maximum/average time interval of $A^{ti}$ & PTIa41\\
& $DA_1^{ti}$ & The difference of $A^{ti}_{max} - A^{ti}_{min}$ & PTIa42\\
& $A^{ti}_{std}$ & The standard deviation of time interval of $A^{ti}$ & PTIa43\\
\noalign{\hrule height 0.5pt}
\end{tabular}
\end{table*}

\begin{table*}[!htbp]
\renewcommand\arraystretch{1.2}
\caption{Combination indicator} 
\label{NotationCI} 
\begin{tabular}{l l l l}
\noalign{\hrule height 0.5pt}
Category & Notation & Description & Dataset\\ 
\noalign{\hrule height 0.5pt}
\textbf{\emph{PAI + PDI}} & $RA_1^{in/out}$ & The ratio of $\frac{A_{{a}_{all}}^{in}}{A_d^{in}}$ and $\frac{A_{{a}_{all}}^{out}}{A_d^{out}}$ & CI1a1\\
& $RA_2$ & The ratio of $\frac{DAa_1}{DAd_1}$ & CI1a2\\
\textbf{\emph{PAI + PTI}} & $Av_a^{in/out}$ &The input/output token amount of an address in each basic unit of the active period & CI2a1\\
& $Av_{a_{avg}}^{in/out}$ & The average input/output token amount of an address in each basic unit of the active period & CI2a11\\
& $Av_{a_{min/max}}^{in/out}$ & The minimum/maximum input/output token amount of an address in each basic unit of the active period & CI2a12\\
& $RAp_2^{in/out}$ & The ratio of  $\frac{Av_a^{in}}{A_p^l}$ and $\frac{Av_a^{out}}{A_p^l}$ & CI2a2\\
& $RAp_{2_{avg}}^{in/out}$ & The average value of $RAp_2^{in/out}$ & CI2a21\\
& $RAp_{2_{min/max}}^{in/out}$ & The minimum/maximum value of $RAp_2^{in/out}$ & CI2a22\\
& $RAp_{2_{std}}^{in/out}$ & The standard deviation of $RAp_2^{in/out}$ & CI2a23\\
& $\Delta A_{a_{all}}^{in/out}$ & The change of $A_{a_{all}}^{in/out}$  \\
& $\Delta RA_1^{in/out}$ & The ratio of $\frac{\Delta A_{a_{all}}^{in}}{A^{ti}}$ and $\frac{\Delta A_{a_{all}}^{out}}{A^{ti}}$ & CI2a3\\
& $\Delta RA_{1_{avg}}^{in/out}$ & The average value of $\Delta RA_1^{in/out}$ & CI2a31\\
& $\Delta RA_{1_{min/max}}^{in/out}$ & The minimum/maximum value of $\Delta RA_1^{in/out}$ & CI2a32\\
& $\Delta RA_{1_{std}}^{in/out}$ & The standard deviation of $\Delta RA_1^{in/out}$ & CI2a33\\
\textbf{\emph{PDI + PTI}} &  $Av_d^{in/out}$ & The in-degree/out-degree of an address in each basic unit of the active period & CI3a1\\
& $Av_{d_{avg}}^{in/out}$ & The average in-degree/out-degree of an address in each basic unit of the active period & CI3a11\\
& $Av_{d_{min/max}}^{in/out}$ & The minimum/maximum in-degree/out-degree of an address in each basic unit of the active period & CI3a12\\
& $RAv_1^{in/out/all}$ & The ratio of  $\frac{Av_d^{in}}{A_{p}^{l}}$, $\frac{Av_d^{out}}{A_{p}^{l}}$, and $\frac{Av_d^{all}}{A_{p}^{l}}$ & CI3a2\\
& $RAv_{1_{avg}}^{in/out/all}$ & The average value of $RAv_1^{in/out/all}$ & CI3a21\\
& $RAv_{1_{min/max}}^{in/out/all}$ & The minimum/maximum value of $RAv_1^{in/out/all}$ & CI3a22\\
& $RAv_{1_{std}}^{in/out/all}$ & The standard deviation of $RAv_1^{in/out/all}$ & CI3a23\\
& $\Delta RA_2^{in/out}$ & The ratio of $\frac{\Delta A_{a_{all}}^{in}}{A^{ti}}$ and $\frac{\Delta A_{a_{all}}^{out}}{A^{ti}}$ & CI3a3\\
& $\Delta RA_{2_{avg}}^{in/out}$ & The average value of $\Delta RA_2^{in/out}$ & CI3a31\\
& $\Delta RA_{2_{min/max}}^{in/out}$ & The minimum/maximum value of $\Delta RA_2^{in/out}$ & CI3a32\\
& $\Delta RA_{2_{std}}^{in/out}$ & The standard deviation of $\Delta RA_2^{in/out}$ & CI3a33\\
\textbf{\emph{PAI + PDI}} & $RA_3^{in}$ & The ratio of $\frac{RA_1^{in}}{A_{p}^{l}}$ & CI4a1 \\
\textbf{\emph{+ PTI}}& $RA_{3_{avg}}^{in}$ & The average value of $RA_3^{in}$ & CI4a11\\
& $RA_{3_{min/max}}^{in}$ & The minimum/maximum value of $RA_3^{in}$ & CI4a12\\
& $RA_{3_{std}}^{in}$ & The standard deviation of $RA_3^{in}$ & CI4a13\\
& $RA_3^{out}$ & The ratio of $\frac{RA_1^{out}}{A_{p}^{l}}$ & CI4a2 \\
& $RA_{3_{avg}}^{out}$ & The average value of $RA_3^{out}$ & CI4a21\\
& $RA_{3_{min/max}}^{out}$ & The minimum/maximum value of $RA_3^{out}$ & CI4a22\\
& $RA_{3_{std}}^{out}$ & The standard deviation of $RA_3^{out}$ & CI4a23\\
& $RA_4^{in}$ & The ratio of $\frac{\Delta RA_2^{in}}{A^{ti}}$ & CI4a3\\
& $RA_{4_{avg}}^{in}$ & The average value of $RA_4^{in}$ & CI4a31\\
& $RA_{4_{min/max}}^{in}$ & The minimum/maximum value of $RA_4^{in}$ & CI4a32\\
& $RA_{4_{std}}^{in}$ & The standard deviation of $RA_4^{in}$ & CI4a33\\
& $RA_4^{out}$ & The ratio of $\frac{\Delta RA_2^{out}}{A^{ti}}$ & CI4a4\\
& $RA_{4_{avg}}^{out}$ & The average value of $RA_4^{out}$ & CI4a41\\
& $RA_{4_{min/max}}^{out}$ & The minimum/maximum value of $RA_4^{out}$ & CI4a42\\
& $RA_{4_{std}}^{out}$ & The standard deviation of $RA_4^{out}$ & CI4a43\\
\noalign{\hrule height 0.5pt}
\end{tabular}
\end{table*}

\subsection{Local Structural Indicator}
Transactions connect addresses in Bitcoin to form a graph, so we believe that considering the characteristics of a subgraph composed of addresses that are close to an address can reflect some information about that address. Unlike the usual $k-hop$ subgraphs, we want to include nodes that are close to each other in the corresponding undirected graph, while reflecting the real network structure. Therefore, we propose the following Algorithm \ref{alg:k-hop} to generate \emph{k}-hop subgraph $G_k$ for each labeled \emph{Ads} node and then obtaining useful \emph{LSI}.

First, we treat the original graph as an undirected graph and traverse each edge in breadth-first order starting from a given address, after which the nodes connected to that edge are renumbered and added to the resulting subgraph in the correct direction by checking the actual direction of that edge in the original graph. In addition, for preventing $G_k$ too large, we set two thresholds for the proposed algorithm. One threshold is the maximum value of \emph{k} and another threshold is the maximum number of \emph{Ads} nodes and \emph{Tx} nodes.

\begin{algorithm}[t]
\caption{$G_k$ generation while preserving original structure} 
\label{alg:k-hop}
\begin{flushleft}
    \textbf{Input:} the dictionary $dict$, the Bitcoin transaction graph $G$, the Bitcoin address $Ads$ \\
    \textbf{Output:} the \emph{k}-hop subgraph $G_k$
\end{flushleft}

\begin{algorithmic}[1]
\Function{Get\_Mapped\_Node}{\var{node}}
    \If{$\var{node} \notin \var{dict}$}
        \State dict(node) $\gets$ len(dict)
    \EndIf
    \State \Return dict(node)
\EndFunction
  
\Function{Gk\_Gen}{\var{G, Ads, depth}}
\State subgraph\_edges $\gets$ list()
 \State Gview $\gets$ \var{G} \textbf{as undirected}
 \State dist(\var{Ads}) $\gets$ 0
 \For{edge in Gview.edges\_in\_bfs\_order(\var{Ads})}
    \State f $\gets$ edge.f
    \State t $\gets$ edge.t
    \If{f $\notin$ dict}
        \State \textbf{break}
    \EndIf
    \If{dist(f) $<$ \var{depth}}
        \If{t $\notin$ dist}
            \State  dist(t) $\gets$ dist(f)+1
        \EndIf
        \State f' $\gets$ Get\_Mapped\_Node(f)
        \State t' $\gets$ Get\_Mapped\_Node(t)
        \If{edge(f,t) $\in$ \var{G.E}}
            \State  subgraph\_edges.append(edge(f',t'))
        \EndIf
        \If{edge(t,f) $\in$ \var{G.E}}
            \State  subgraph\_edges.append(edge(t',f'))
        \EndIf
    \EndIf
\EndFor
\State \Return $G_k$ 
\EndFunction
\end{algorithmic}
\end{algorithm}

It is noted that in our implemented Algorithm \ref{alg:k-hop}, for the first threshold, we choose 3,000 as the maximum number of the sum of \emph{Ads} node and \emph{Tx} node, and for another one, we select $k=4$ as the maximum value. We will explain the reason below using Fig. \ref{BTC_Gk}. 

In Fig. \ref{BTC_Gk}\footnote{Using $Ads$ node \emph{1BtcBoSSnqe8mFJCUEyCNmo3EcF8Yzhpnc} from Ponzi}, there are 8 figures of $G_k$ with different numbers of nodes generated from the same $Ads$, from which it is clear that Fig. \ref{n0}, Fig. \ref{n1}, and Fig. \ref{n2} contain less structural information than other 5 figures of $G_k$. Besides, Fig. \ref{n6} and Fig. \ref{n7} have too much information that is similar to the entire transaction graph that can not easily get valuable distinctive information. Compared among the rest Fig. \ref{n3}, Fig. \ref{n4}, and Fig. \ref{n5} we observe that their structural are almost identical, thus, we choose Fig. \ref{n3} (i.e., $G_k$ with 3,000 nodes) considering the efficiency of generate $G_k$. It is noted that we find similar conclusions in many other address nodes in different types of Bitcoin addresses during our plenty of experiments. Also, through the experiments, we find this set of hyperparameters achieves a balance between our device's performance limit and the effect of indicators from the local Bitcoin transaction graph. The specific features are shown below: 

\begin{figure*}[!htbp]
\centering
\subfigure[500 nodes]{
\includegraphics[width=0.2\linewidth]{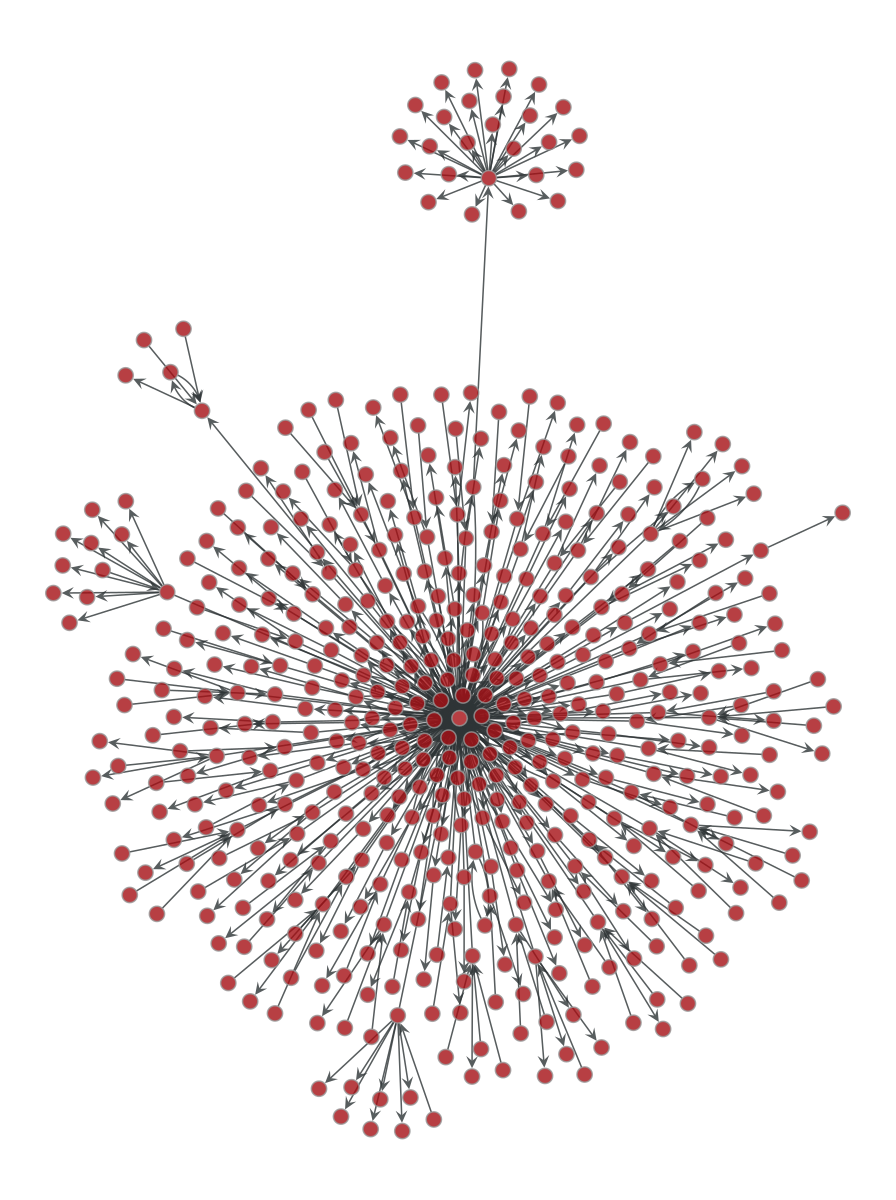}
\label{n0}
}
\quad
\subfigure[1000 nodes]{
\includegraphics[width=0.2\linewidth]{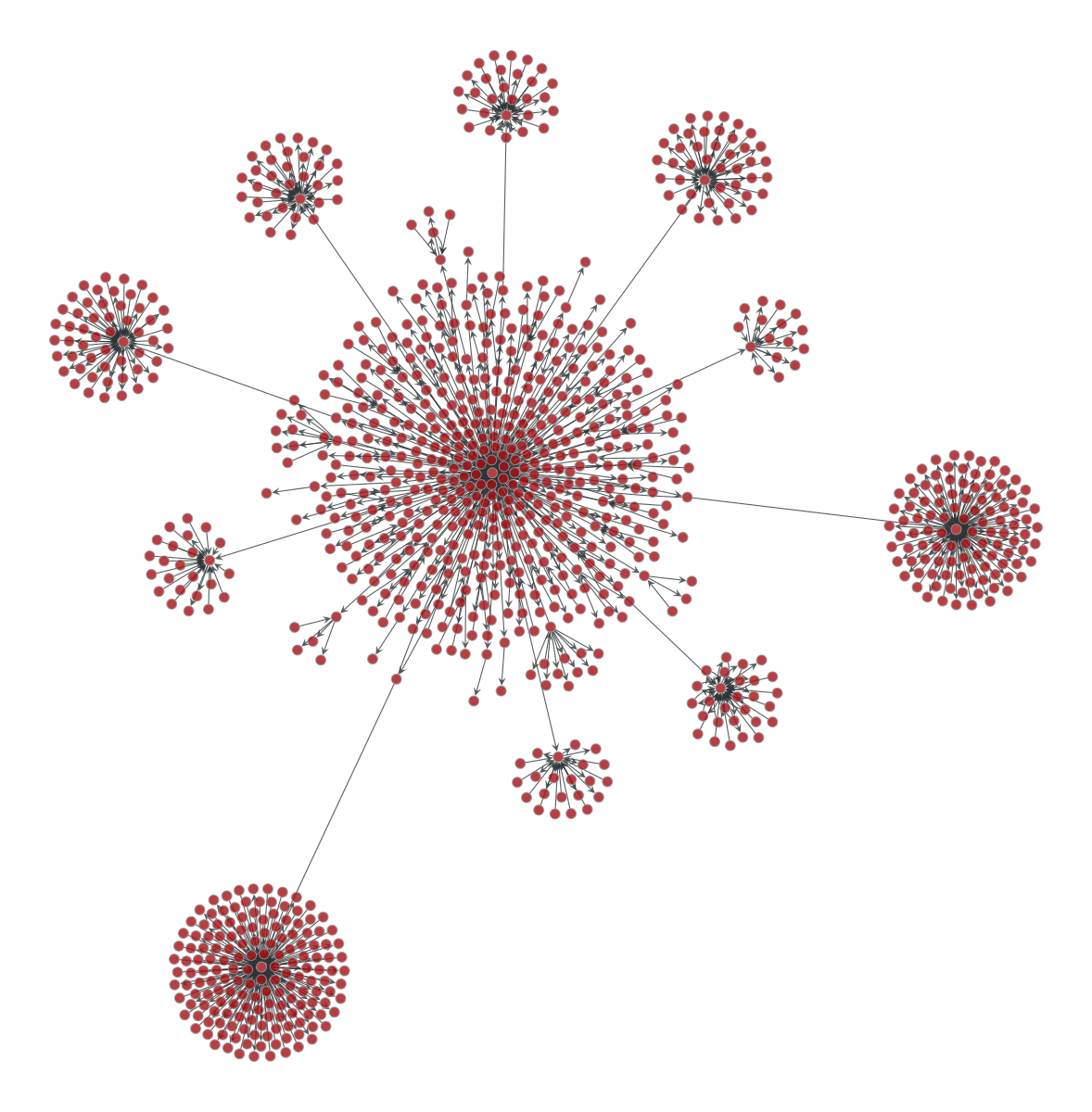}
\label{n1}
}
\quad
\subfigure[2000 nodes]{
\includegraphics[width=0.2\linewidth]{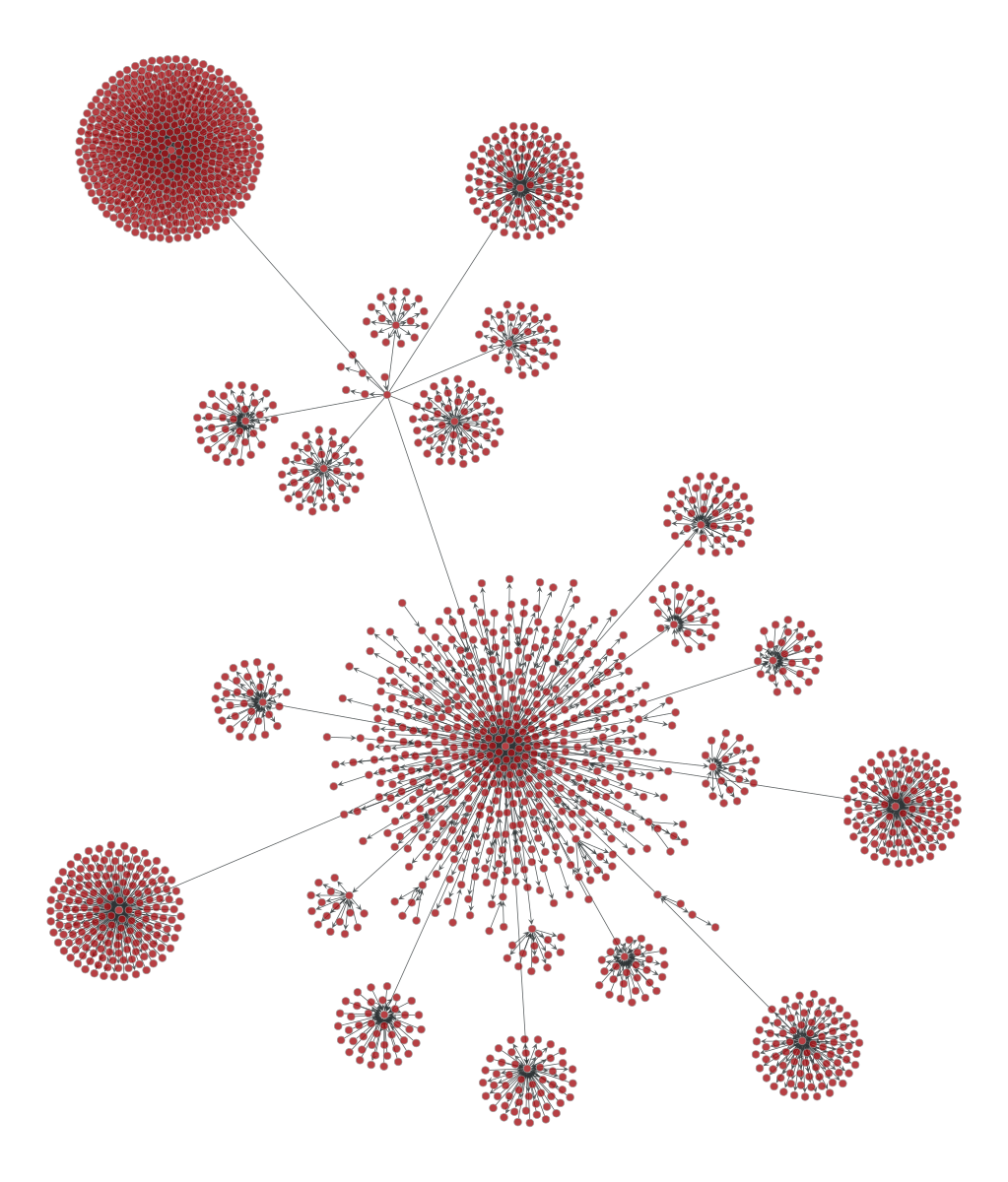}
\label{n2}
}
\quad
\subfigure[3000 nodes]{
\includegraphics[width=0.2\linewidth]{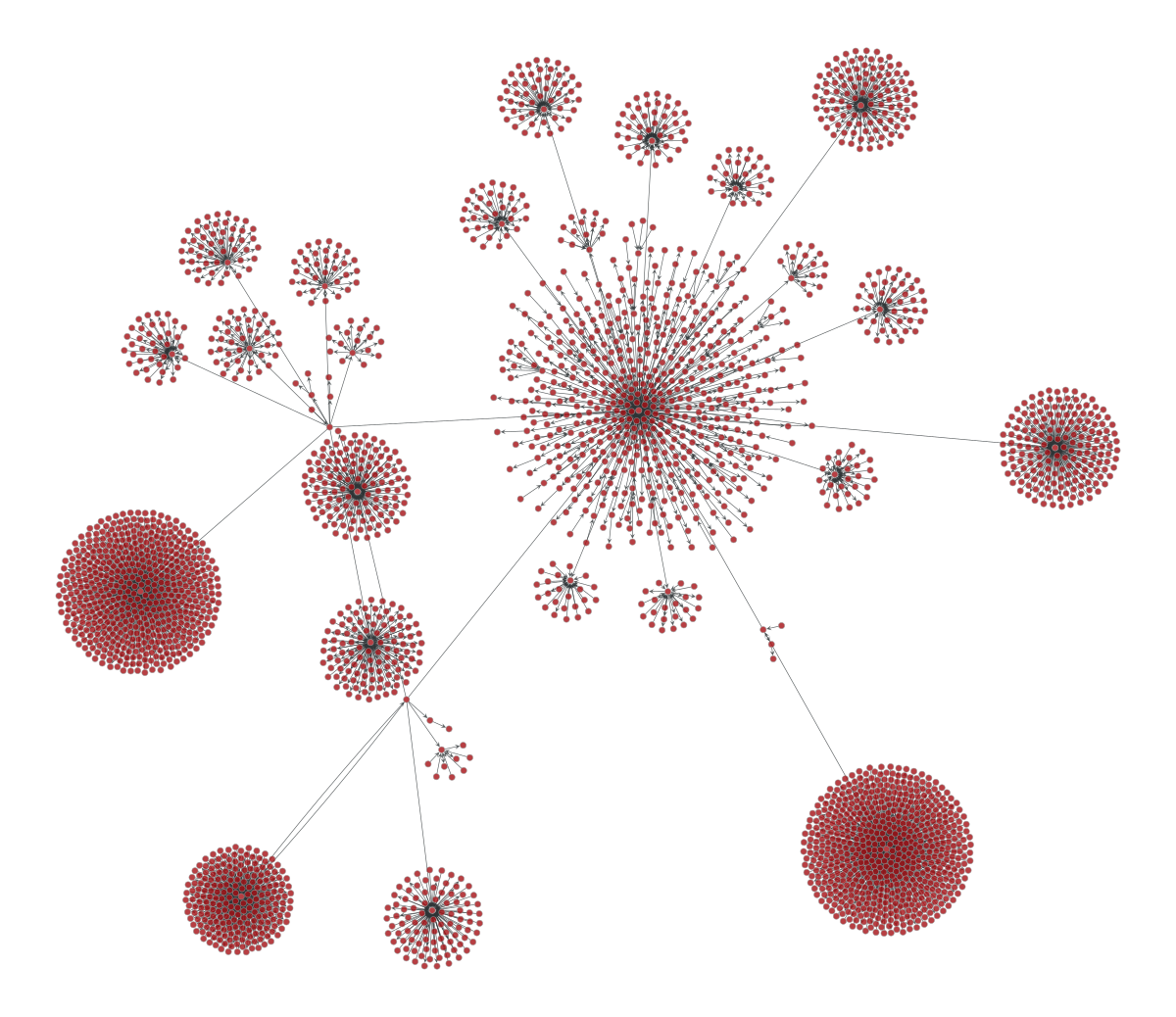}
\label{n3}
}
\quad
\subfigure[4000 nodes]{
\includegraphics[width=0.2\linewidth]{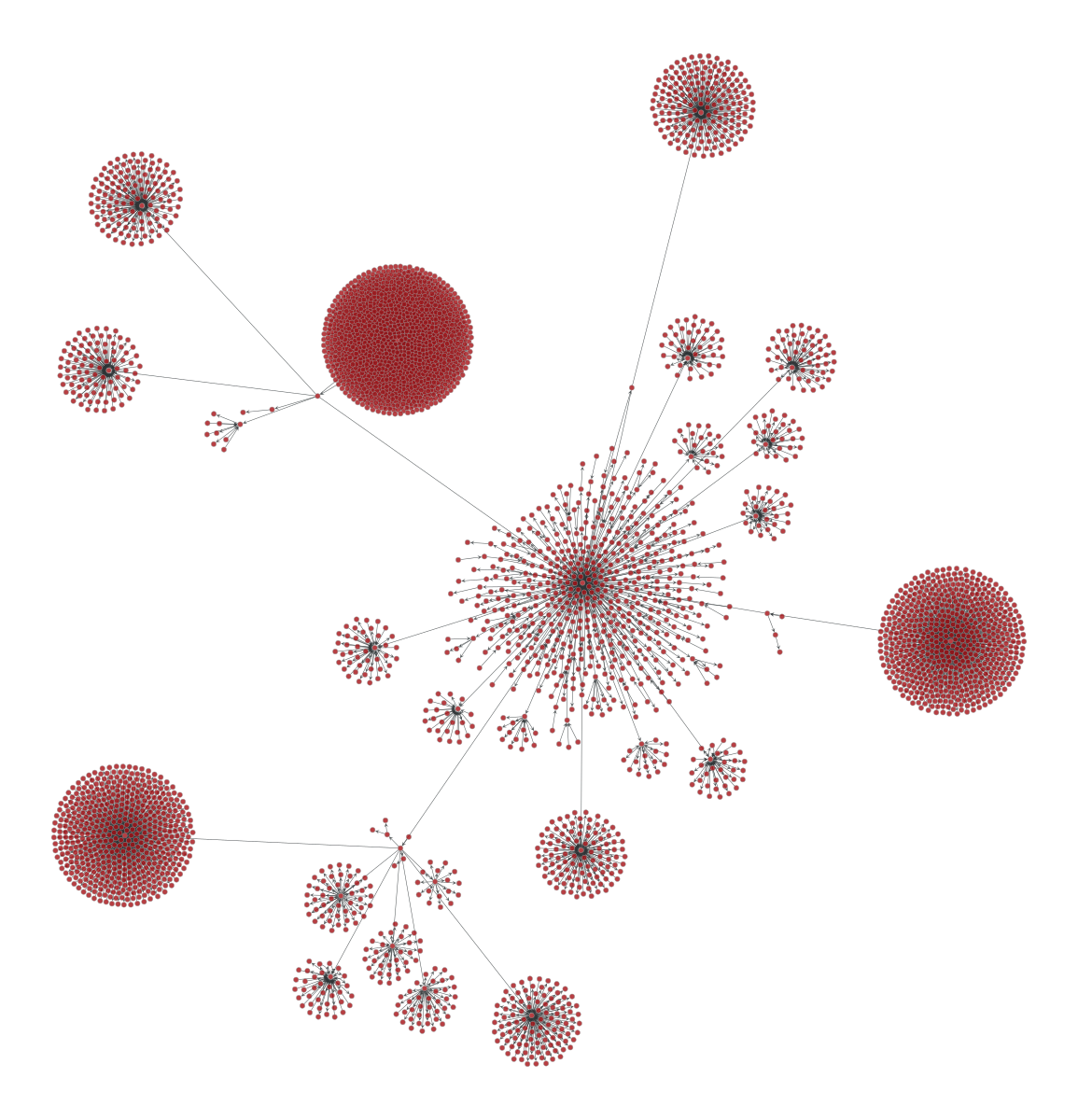}
\label{n4}
}
\quad
\subfigure[5000 nodes]{
\includegraphics[width=0.2\linewidth]{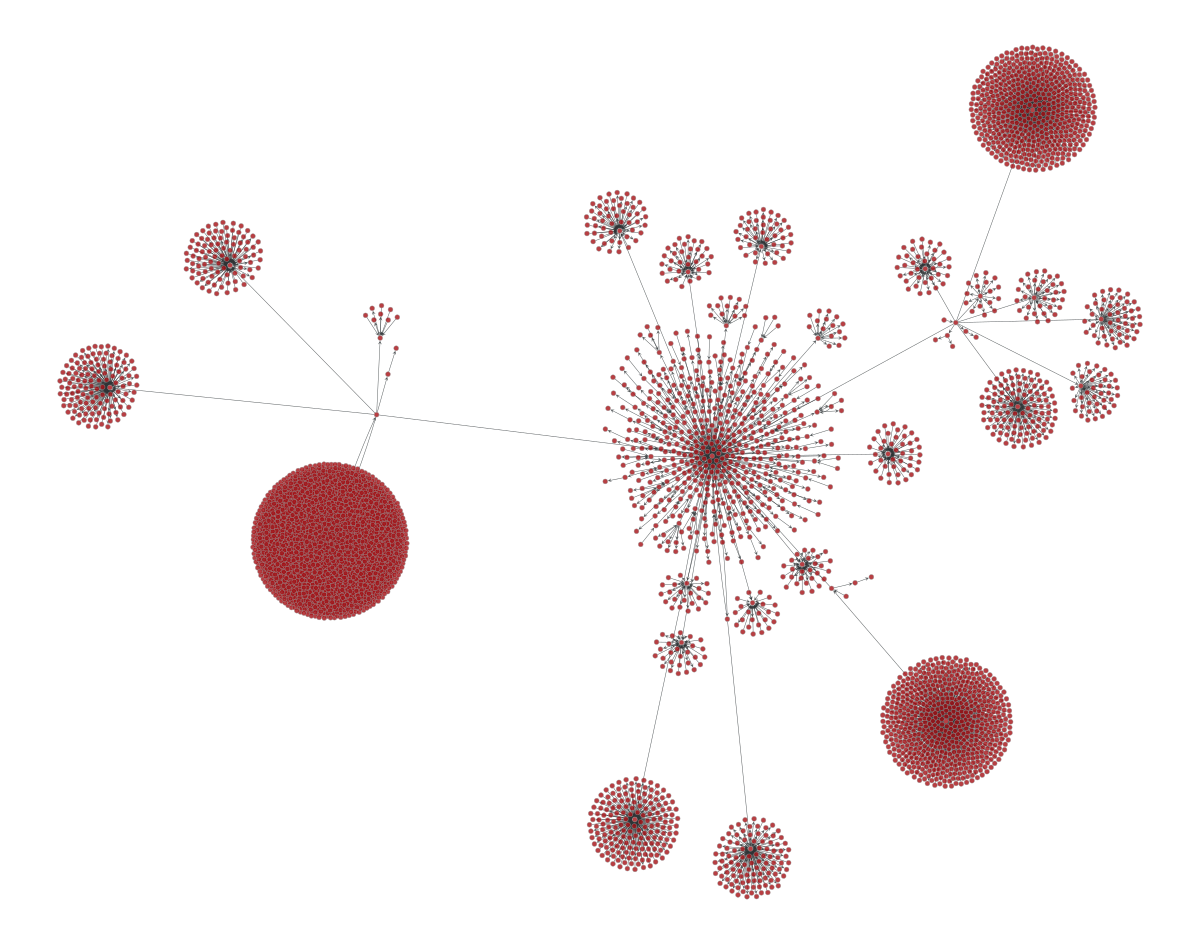}
\label{n5}
}
\quad
\subfigure[10000 nodes]{
\includegraphics[width=0.2\linewidth]{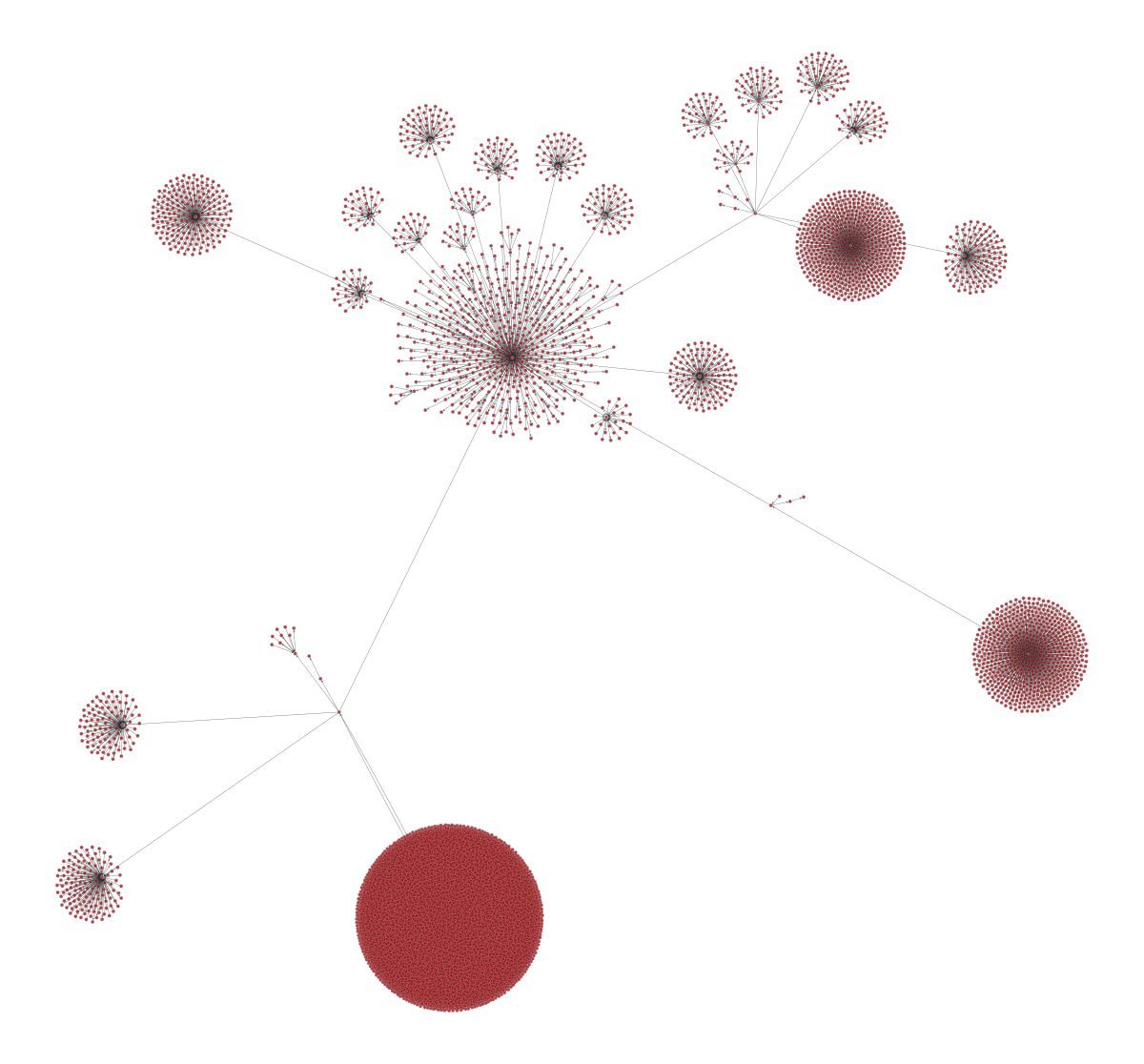}
\label{n6}
}
\quad
\subfigure[20000 nodes]{
\includegraphics[width=0.2\linewidth]{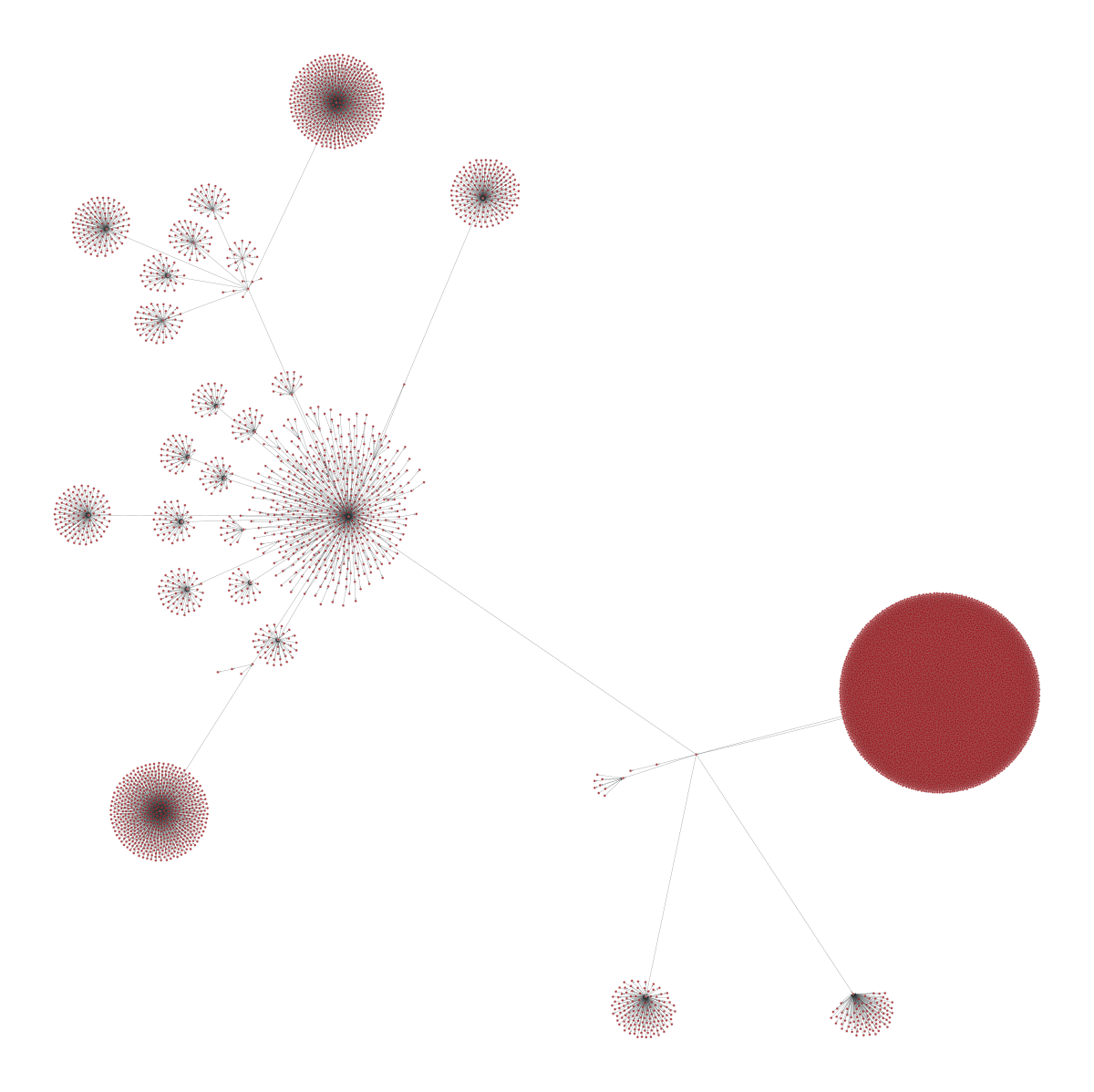}
\label{n7}
}
\caption{$G_k$ with maximum $k=4$ and different maximum number of nodes}
\label{BTC_Gk}
\end{figure*}

\begin{itemize}
    \item \textbf{Average Degree (S1).}
    $G_k^{\overline d}$ means the average degree \cite{Boccaletti1} of the nodes in $G_k$:
    \begin{equation}
        G_k^{\overline d} = \frac{1}{G_k^A+G_k^T}(\sum_{i=1}^{G_k^A}A_i+\sum_{j=1}^{G_k^T}T_j),
    \end{equation}
    where $G_k^A$ is the number of \emph{Ads} nodes, $G_k^T$ is the number of \emph{Tx} nodes, $A_i$ is the degree of the \emph{Ads} node, and $T_j$ is the degree of the \emph{Tx} node. Specifically, we calculate the average values and standard deviations of in-degree/out-degree/in-degree and out-degree in $G_k$ as features in \emph{LSI}.
    
    \item \textbf{Degree Distribution (S2).} 
    $G_k^{D(d)}$ is the degree distribution \cite{Boccaletti1} that means the proportion of the node with the \emph{d} degree in $G_k$ \cite{Boccaletti1}. Specifically, the degree here can be divided into three types that are in-degree, out-degree, and in-degree and out-degree. Besides, we select the maximum value of degree distribution as the feature of \emph{LSI}.
    
    \item \textbf{Degree Correlation (S3).} ${G_k^{C_d}}$ is the degree correlation \cite{Boccaletti1} of $G_k$ that describes the relationship between nodes with large degree and nodes with small degree. In this paper, simplified Pearson degree correlation proposed by Newman \cite{Newman} is selected to measure $G_k$:
    \begin{equation}
    G_k^{C_d} = \frac{\frac{1}{E}\sum_{e_{ij}}d_id_j-[\frac{1}{E}\sum_{e_{ij}}\frac{1}{2}(d_i+d_j)]^2}{\frac{1}{E}\sum_{e_{ij}}{(d_i^2+d_j^2)}-[\frac{1}{E}\sum_{e_{ij}}\frac{1}{2}(d_i+d_j)]^2},
    \end{equation}
    where the edge $e_{ij}$ connects $node_i$ and $node_j$ then $d_i$ and $d_j$ stand for these two nodes' degree respectively, $E$ is the total number of edges. 
    
    \item \textbf{Betweenness (S4).} $G_k^{B_i}$ is the betweenness that reflects the impact of the node in $G_k$ \cite{Boccaletti1}:
    \begin{equation}
        G_k^{B_i} = \sum_{i \neq j \neq k}\frac{P_{jk}^s(i)}{P_{jk}^s},
    \end{equation}
    where $P_{jk}^s$ is the shortest path between $node_j$ and $node_k$, $P_{jk}^s(i)$ is the number of shortest path between $node_j$ and $node_k$ while going through $node_i$.
    
    \item \textbf{Average Path (S5).} $G_k^{\overline P}$ is the average distance between any two nodes in $G_k$:
    \begin{equation}
        G_k^{\overline P} = \frac{1}{(G_k^A+G_k^T)(G_k^A+G_k^T-1)}\sum_{i \neq j}len_{ij},
    \end{equation}
    where $len_{ij}$ is the distance between $node_i$ and $node_j$.
    
    \item \textbf{Diameter (S6).} $G_k^{len}$ is the longest distance between any two nodes in $G_k$.
    

    \item \textbf{Closeness Centrality (S7).} $G_k^{Ce_i}$ measures the closeness from one node to other nodes, which can provide the efficiency of spreading information of a node \cite{okamoto}:
    \begin{equation}
        G_k^{Ce_i} = \frac{1}{\sum_{j=1}^{n-1}len_{min}(i,j)},
    \end{equation}
    where $n$ is the total number of nodes in $G_k$, $len_{min}(i,j)$ is the minimum distance between $node_i$ and $node_j$, and $node_i$ is the node required to be measured.
    
    \item \textbf{PageRank (S8).} $G_k^{PR(i)}$ is value of PageRank of $node_i$ in $G_k$, which evaluates the influence of the node \cite{Lawrence}:
    \begin{equation}
    G_k^{PR(i)}=\frac{1-\alpha}{G_k^A+G_k^T}+\alpha \sum_{j \in \beta(i)}\frac{G_k^{PR(j)}}{\gamma(j)},
    \end{equation}
    where $\alpha$ is a damping factor, $\beta(i) $ are the in-neighbors of $node_i$, and $\gamma(j) $ is the out-degree of $node_j$. 
    
    \item \textbf{Density (S9).} $G_k^{ds}$ is the density of $G_k$ that measures the density according to the edge connectivity \cite{bollobas1998modern}:
    \begin{equation}
        G_k^{ds} = \frac{E}{(G_k^A+G_K^T)(G_K^A+G_K^T-1)},
    \end{equation}
    where $E$ is the total number of edges.
\end{itemize}

\section{Experiments}
\subsection{Basic Setting}
We mainly utilize \emph{graph-tool} \cite{peixoto_graph-tool_2014} to implement the framework proposed above. Also, we use \emph{networkx} \cite{Hagberg} to test and verify our methods and algorithms on small graphs. Our device configuration is shown in Table \ref{device} and our code can be found on GitHub\footnote{https://github.com/Y-Xiang-hub/Bitcoin-Address-Behavior-Analysis}.

\begin{table}[!htbp]
    \centering
    \caption{Computer Configuration}
    \label{device}
    \begin{tabular}{*2l}
        \toprule
        Hardware / Software &  Configuration \\
        \hline
        \midrule
        CPU & Intel Xeon Silver 4210R 2.40 GHz\\
        RAM & 320 GB \\
        Operating System & Ubuntu 21.04 \\
        Python & 3.9.5 \\
        graph-tool & 2.42 \\
        \bottomrule
    \end{tabular}
\end{table}

Additionally, we mainly use \emph{scikit-learn} \cite{scikit-learn} and \emph{xgboost} \cite{xgboost} to implement common machine learning models. We choose \emph{k}-nearest neighbors (KNN) algorithm, decision tree (DT), random forest (RF), multilayer perceptron (MLP), and XGBoost (XGB) to test the BABD-13.

\subsection{Experiment Process}
After the Bitcoin ledger (JSON files) and labeled Bitcoin both have been collected completely, our following experiment consists of three phases - transaction graph construction, features extraction, and address category modeling. The specific experiment implementation is illustrated below.

First, we construct the directed heterogeneous multigraph-based Bitcoin transaction graph through JSON files of Bitcoin ledger as Fig. \ref{BTC_structure} by graph generation function implemented by \emph{graph-tool} (the JSON example and code can be found in our project). The point of this step is that coinbase transactions are different from normal transactions, thus, we add them to the graph in different methods respectively.

Then, we extract intuitive features in \emph{SI} of labeled Bitcoin addresses directly from the constructed Bitcoin graph, leveraging functions implemented by \emph{graph-tool}. However, it is more complicated and slower for extracting features from \emph{LSI} because we need to use Algorithm \ref{alg:k-hop} to generate a concrete subgraph for each labeled Bitcoin address and then obtain the features by complex network functions provided by \emph{graph-tool} from the generated $G_k$. 

Besides, we adopt parallel computing ways to accelerate the speed to get the features of Bitcoin addresses. Next, we store the result in a CSV file and preprocess it. There is an important step in preprocessing the raw CSV file, which is filling the missing values. To deal with it, we use the zero value to fill all the missing values in the raw CSV file according to our understanding of our proposed features on the Bitcoin transaction graph.

Finally, we use \emph{scikit-learn} and \emph{xgboost} to implement common machine learning models to test BABD-13\footnote{For easily reproducing our work, we set parameter \emph{random\_state=9} for every method and model if it exists}. We perform an 8:2 split of the training set and testing set on the BABD-13 using function $train\_test\_split()$ with parameter \emph{test\_size=0.2}. Then we choose the min-max normalization to preprocess the split BABD-13 training set and testing set by function $MinMaxScaler()$. It is noted that we divide the BABD-13 into two parts that are \emph{SI} with 132 features and \emph{LSI} with 16 features. 

The task for these models is a multiclass classification on BABD-13 and the number of samples for each class is quite imbalanced. We select 5 machine learning models to test BABD-13. The models with their corresponding methods in \emph{scikit-learn} and parameter configurations are shown below.

\begin{itemize}
    \item \textbf{KNN}, built by function $KNeighborsClassifier()$ from \emph{scikit-learn} with parameters  \emph{n\_neighbors=4}, \emph{algorithm=`kd\_tree'}, and \emph{weights=`distance'}.
    \item \textbf{DT}, built by function $DecisionTreeClassifier()$ from \emph{scikit-learn} with parameters \emph{criterion=`entropy'} and \emph{splitter=`best'}.
    \item \textbf{RF}, built by function $RandomForestClassifier()$  with parameter \emph{n\_estimators=200}.
    \item \textbf{MLP} by function $MLPClassifier()$ from \emph{scikit-learn} with parameters \emph{max\_iter=1000}, \emph{solver=`adam'}, and \\ \emph{hidden\_layer\_sizes=(100, 100, 100)}.
    \item \textbf{XGB} by function $XGBClassifier()$ from \emph{xgboost} with parameters \emph{objective=`multi:softmax'}, \emph{num\_class=13}, \\ \emph{eval\_metric=`mlogloss'}, \emph{learning\_rate=0.5}.
\end{itemize}

\subsection{Result and Analysis}
The experiment results for the 13-classification task of different 5 machine learning models are shown in Table \ref{ml}\footnote{The parameter \emph{average=`weighted'} is used for calculating precision, recall, and f1-score}. The results of (\emph{SI+LSI}) witness that the accuracy rate, precision, recall, and f1-score can reach at least 93.24\%, 92.80\%, 93.24\%, and 92.97\%, and at most 96.71\%, 96.46\%, 96.71\%, and 96.57\%.

\begin{table}[!htbp]
\caption{The Performance of Different Machine Learning Models}
\label{ml}
\centering
\begin{tabular}{l cccc}
\toprule
\multirow{2}{*}{Method} 
        & \multicolumn{4}{c}{Evaluation Metric} \\
\cmidrule(lr){2-5} 
        & Accuracy & Precision & Recall & F1-score \\
\hline
\midrule
${\rm KNN}^{SI}$ & 0.6862 & 0.6667 & 0.6862 & 0.6749 \\
\addlinespace
${\rm KNN}^{LSI}$ & 0.9418 & 0.9381 & 0.9418 & 0.9397 \\
\addlinespace
${\rm KNN}^{SI+LSI}$ & 0.9324 & 0.9280 & 0.9324 & 0.9297 \\
\addlinespace
${\rm DT}^{SI}$ & 0.7061 & 0.6961 & 0.7061 & 0.7001 \\
\addlinespace
${\rm DT}^{LSI}$ & 0.9376 & 0.9375 & 0.9376 & 0.9375 \\
\addlinespace
${\rm DT}^{SI+LSI}$ & 0.9444 & 0.9447 & 0.9444 & 0.9446 \\
\addlinespace
${\rm RF}^{SI}$& 0.7566 & 0.7366 & 0.7566 & 0.7342 \\
\addlinespace
${\rm RF}^{LSI}$ & 0.9565 & 0.9532 & 0.9565 & 0.9544 \\
\addlinespace
${\rm RF}^{SI+LSI}$ & 0.9598 & 0.9564 & 0.9598 & 0.9577 \\
\addlinespace 
${\rm MLP}^{SI}$ & 0.7366 & 0.7081 & 0.7366 & 0.7063 \\
\addlinespace 
${\rm MLP}^{LSI}$ & 0.9331 & 0.9289 & 0.9331 & 0.9306 \\
\addlinespace 
${\rm MLP}^{SI+LSI}$ & 0.9406 & 0.9376 & 0.9406 & 0.9385 \\
\addlinespace 
${\rm XGB}^{SI}$ & 0.7825 & 0.7624 & 0.7825 & 0.7654 \\
\addlinespace 
${\rm XGB}^{LSI}$ & 0.9604 & 0.9579 & 0.9604 & 0.9590 \\
\addlinespace 
${\rm XGB}^{SI+LSI}$ & 0.9671 & 0.9646 & 0.9671 & 0.9657 \\
\addlinespace 
\bottomrule
\end{tabular}
\end{table}

From the results, it is also obvious that for the two kinds of indicators \emph{LSI} and \emph{SI}, the former performs better than \emph{SI} greatly. The average accuracy rates of these 5 models are 94.58\% and 73.36\% respectively on \emph{LSI} and \emph{SI}. Also, for example, on KNN, the accuracy rate improves by 25.56\% after using \emph{LSI} instead of \emph{SI} which is the maximal increase among these models. Besides, at least the accuracy rate enhances 17.79\% on XGB after applying \emph{LSI} instead of \emph{SI}.

However, the influence of \emph{SI} can not be ignored because we can see that the combination of \emph{SI+LSI} leads to the increment of all evaluation metrics to a certain extent for all the models except KNN. Although the improvement is not significant, it proves that there are valuable features in \emph{SI} and we will further analyze them in the next section.

Additionally, from the perspective of the model level, XGB displays the best results that are all higher than 96.46\% on all evaluation metrics and RF is following whose results are all higher than 95.64\%. The results of evaluation metrics show that each metric of DT and MLP is close, and the differences in each metric are fewer than 0.71\%. Besides, the differences in each metric for MLP and KNN range from 0.82\% to 0.96\%. Besides, MLP performs better clearly when using only \emph{SI} compared with the other two models, the accuracy rate is 73.66\% and is close to RF.

\begin{figure*}[!htbp]
  \centering
  \includegraphics[width=\linewidth]{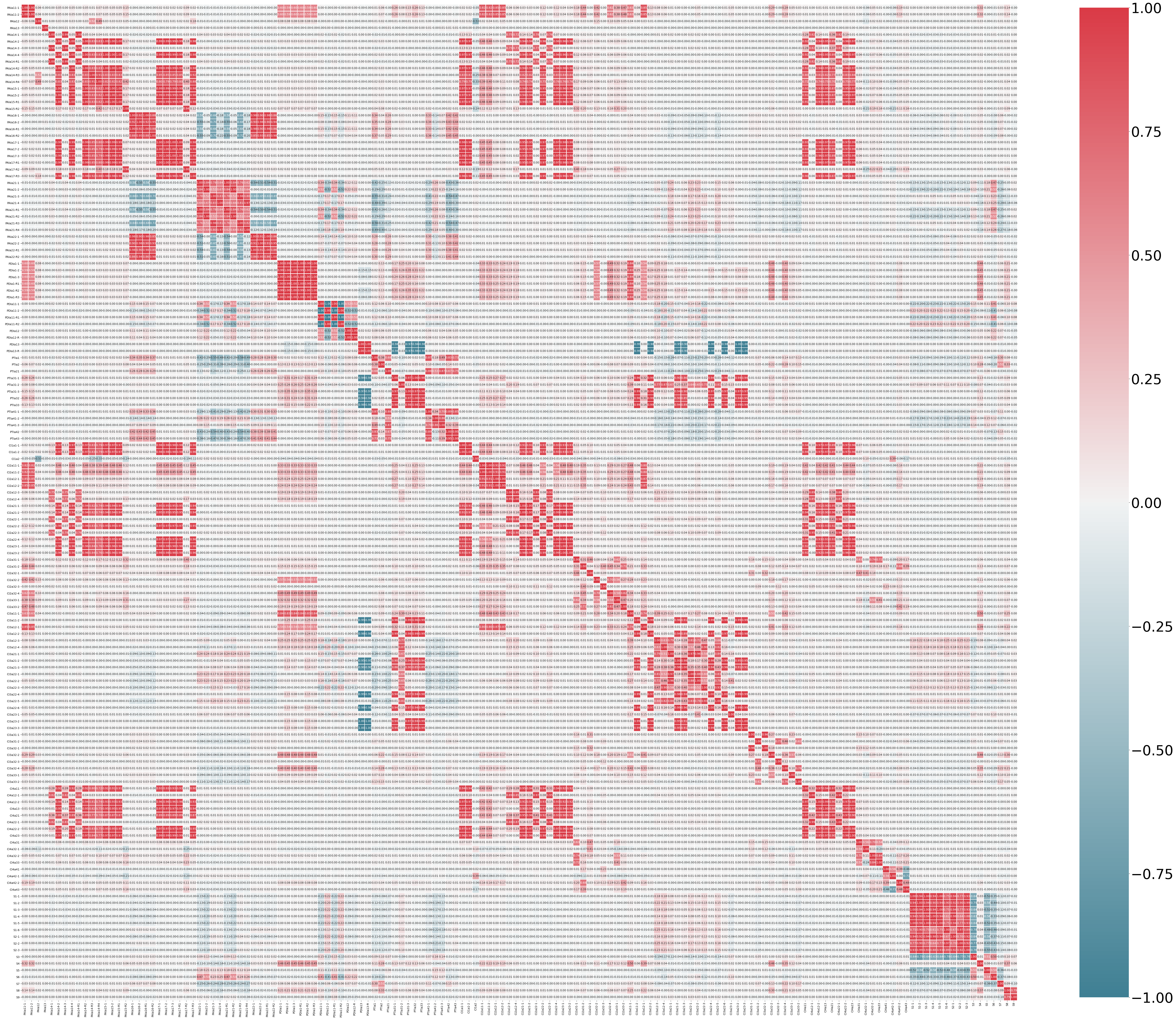}
  \caption{Bitcoin address feature heatmap}
  \label{heatmap}
\end{figure*}

According to the evaluation metrics, the effect order of these 5 models on BABD-13 is XGB$>$RF$>$DT$\approx$MLP$>$KNN on BABD-13. Except for the above analysis, we also take a closer look at the relationships among these 148 features on BABD-13 using a heatmap, from which we can observe the degree of relevance between every two features, and we will use the heatmap to analyze features in BABD-13 in the following section. The generated heatmap of BABD-13 is shown in Fig. \ref{heatmap}. 

To sum up, these 5 common machine learning models show good performance on BABD-13, and it proves that BABD-13 is meaningful and helpful for further studying the behavioral patterns of different types of Bitcoin addresses. Moreover, some specific features that are listed on BABD-13 need to be considered further for the Bitcoin address behaviors study. Besides, we compared our scheme with other similar schemes, the results can be seen below in Table \ref{compare}.

\begin{table*}[!hbtp]
\caption{Comparison with other schemes}
\label{compare}
\centering
\begin{tabular}{|c|c|c|c|c|c|c|}
\hline
\textbf{\diagbox{Comparison}{Scheme}} 
& \makecell*[c]{\textbf{BABD}}
& \makecell*[c]{\textbf{\cite{Weber}}} 
& \makecell*[c]{\textbf{\cite{Li}}}
& \makecell*[c]{\textbf{\cite{Ranshous}}}
& \makecell*[c]{\textbf{\cite{michalski2020revealing}}}
& \makecell*[c]{\textbf{\cite{8751410}}}
\\[1ex] \hline

\emph{\makecell{Time range}} 
& \makecell*[c]{Jul. 12, 2019 \\ $\sim$  \\ May 26, 2021}    
& \textbackslash  
& \textbackslash  
& \makecell*[c]{Sep. 29, 2011 \\ $\sim$ \\ Apr. 22, 2015}  
& \makecell*[c]{May 2, 2018 \\ $\sim$ \\ May 3, 2018} 
& \makecell*[c]{Jan. 3, 2009 \\ $\sim$ \\ Jun. 30, 2018}   
\\ \hline

\emph{Graph size}        
& \makecell*[c]{516,167,131 nodes \\ 713,703,239 edges}
& \makecell*[c]{203,769 nodes \\ 234,355 edges}  
& \textbackslash 
& \textbackslash 
& \textbackslash 
& \textbackslash 
\\ \hline

\makecell*[c]{\emph{Sample size}}   
& \makecell*[c]{544,462}        
& \makecell*[c]{46,564}
& \makecell*[c]{1,234,047}
& \makecell*[c]{972,866}
& \makecell*[c]{8,808}
& \makecell*[c]{26,313}
\\[1ex] \hline

\makecell*[c]{\emph{Object}}            
& \makecell*[c]{\emph{Ads} node}
& \makecell*[c]{\emph{Tx} node}
& \makecell*[c]{\emph{Ads} node}
& \makecell*[c]{\emph{Ads} node}
& \makecell*[c]{\emph{Ads} node}
& \makecell*[c]{\emph{Ads} node}
\\[1ex] \hline

\makecell*[c]{\emph{Feature size}}    
& \makecell*[c]{148}                  
& \makecell*[c]{166}
& \makecell*[c]{92}
& \makecell*[c]{\textbackslash}
& \makecell*[c]{149}
& \makecell*[c]{64}
\\[1ex] \hline

\makecell*[c]{\emph{Number of types}}
& \makecell*[c]{13}                  
& \makecell*[c]{2}
& \makecell*[c]{2}
& \makecell*[c]{2}
& \makecell*[c]{6}
& \makecell*[c]{7}   
\\[1ex] \hline

\emph{Method}           
& \makecell*[c]{DT, KNN, MLP, \\ RF, XGB}         
& \makecell*[c]{GCN, LR, MLP, \\ RF}
& \makecell*[c]{ANN, RF, SVM, \\ XGB }
& \makecell*[c]{AdaBoost, LR, \\ MLP, RF, SVM}
& \makecell*[c]{DT, ET, NN, \\ RF, SVM}
& \makecell*[c]{AdaBoost, LightGB, \\ LR, MLP, NN,\\ RF, SVM, XGB }
\\ \hline

\emph{Best performance}       
& \makecell*[l]{Accuracy: 97.1\% \\ Precision: 96.9\% \\ Recall: 97.1\% \\ F1-score: 97.0\%}                 
& \makecell*[l]{Precision: 97.1\% \\ Recall: 66.80\% \\ F1-score: 98.6\%}
& \makecell*[l]{Precision: 93.5\% \\ Recall: 85.4\% \\ F1-score: 88.1\%}
& \makecell*[l]{ Precision: 99.7\% \\ Recall: 99.7\% \\ F1-score: 99.7\%}
& \makecell*[l]{F1-score: 96.0\% }
& \makecell*[l]{F1-score: 87.0\% } 
\\ \hline

\makecell*[c]{\emph{Dataset available}}
& \makecell*[c]{\cmark}
& \makecell*[c]{\cmark}
& \makecell*[c]{\cmark}
& \makecell*[c]{\xmark}
& \makecell*[c]{\cmark}
& \makecell*[c]{\xmark}
\\[1ex] \hline

\makecell*[c]{\emph{Code available}}
& \makecell*[c]{\cmark}
& \makecell*[c]{\xmark}
& \makecell*[c]{\xmark}
& \makecell*[c]{\xmark}
& \makecell*[c]{\xmark}
& \makecell*[c]{\xmark}
\\[1ex] \hline

\makecell*[c]{\emph{Year}}      
& \makecell*[c]{2022}        
& \makecell*[c]{2019}
& \makecell*[c]{2020}
& \makecell*[c]{2017}
& \makecell*[c]{2020}
& \makecell*[c]{2019}
\\[1ex] \hline

\end{tabular}
\end{table*}

\subsection{Feature Selection}
\label{sec:D-FS}
We utilize two methods to measure the importance of each feature in BABD-13, one is the function $SelectKBest()$ from \emph{scikit-learn}, and another is the attribute \emph{feature\_importances\_}. The first method is used for general single feature selection and the second one is applied for a specific model to choose features.

Based on the feature selection results on both ways, we find that the top 10 features are listed as follows:
\begin{enumerate}
\item The maximum out-degree in $G_k$ (S2-2).
\item The standard deviation of the in-degree and out-degree in $G_k$ (S1-6). 
\item The standard deviation of in-degree in $G_k$ (S1-2). 
\item The degree correlation of $G_k$ (S3). 
\item $RAa_{3_{min}}^{in}$ (PAIa21-1). 
\item $A_{min}^{ti}$ (PTIa41-2). 
\item The longest distance between any two nodes in $G_k$ (S6). 
\item The closeness centrality of $G_k$ (S5).
\item $\Delta RA_{2_{max}}^{in}$ (CI3a32-2). 
\item The density of $G_k$ (S7). 
\end{enumerate}

In addition, we also rank the top 1 feature of each indicator type (i.e., \emph{PAI}, \emph{PDI}, \emph{PTI}, \emph{CI}, and \emph{LSI}) are showed below:
\begin{itemize}
\item \emph{PAI}: $RAa_{3_{min}}^{in}$ (PAIa21-1). 
\item \emph{PDI}: The out-degree of an address (PDIa1-2).
\item \emph{PTI}: $A_{min}^{ti}$ (PTIa41-2). 
\item \emph{CI}: $\Delta RA_{2_{max}}^{in}$ (CI3a32-2). 
\item \emph{LSI}: The maximum out-degree in $G_k$ (S2-2).
\end{itemize}

\subsection{Model Improvement by Selected Features}

The final results on KNN and RF the new selected feature set are shown in Table \ref{ml-2}. It is obvious that after further selecting features in \emph{SI} on BABD-13, the performances of KNN and RF improve. Specifically, the accuracy rate increases by 1.20\% on KNN which is obvious. Additionally, the accuracy rates on RF and XGB go up by 0.37\% and 0.42\% which are not evident compared to KNN. But the XGB reaches the new highest level of accuracy rate of 97.13\% after utilizing crafted chosen features. As for other evaluation metrics in Table \ref{ml-2} for KNN and RF, there are similar increments as the corresponding accuracy rate trends above.

The experimental results prove that the current feature combination, i.e., \emph{SI+LSI} can be re-selected in \emph{SI}, for higher model performance for different machine learning models. However, for different models, the same feature might have an unequal influence (positive or negative) that will cost time to test a number of feature combinations to obtain more results. The effect on different machine learning models due to feature selection on BABD-13 can be further studied in future work. The experiment records are stored in our GitHub project.

\begin{table}[!htbp]
\caption{The Performance on KNN, RF, and XGB Using Selected new Features}
\label{ml-2}
\centering
\begin{tabular}{l cccc}
\toprule
\multirow{2}{*}{Method} 
        & \multicolumn{4}{c}{Evaluation Metric} \\
\cmidrule(lr){2-5} 
        & Accuracy & Precision & Recall & F1-score \\
\hline
\midrule
\addlinespace
${\rm KNN}^{SI+LSI}$ & 0.9324 & 0.9280 & 0.9324 & 0.9297 \\
\addlinespace
$\textbf{KNN}^\textbf{new}$ & \textbf{0.9444} & \textbf{0.9409} & \textbf{0.9444} & \textbf{0.9424} \\
\midrule
\addlinespace
${\rm RF}^{SI+LSI}$ & 0.9598 & 0.9564 & 0.9598 & 0.9577 \\
\addlinespace
$\textbf{RF}^\textbf{new}$ & \textbf{0.9635} & \textbf{0.9603} & \textbf{0.9635} & \textbf{0.9616} \\
\midrule
\addlinespace 
${\rm XGB}^{SI+LSI}$ & 0.9671 & 0.9646 & 0.9671 & 0.9657 \\
\addlinespace
$\textbf{XGB}^\textbf{new}$ & \textbf{0.9713} & \textbf{0.9693} & \textbf{0.9713} & \textbf{0.9702} \\
\bottomrule
\end{tabular}
\end{table}

\subsection{Feature Analysis}
In this part, we select the top 1 feature in each type of indicator mentioned in Section \ref{sec:D-FS} and analyze the effect and data distribution of features in BABD-13. 

\begin{figure*}[!htbp]
\centering
\subfigure[PAIa21-1]{
\begin{minipage}[t]{0.3\linewidth}
\label{s1}
\centering
\includegraphics[width=1.75in]{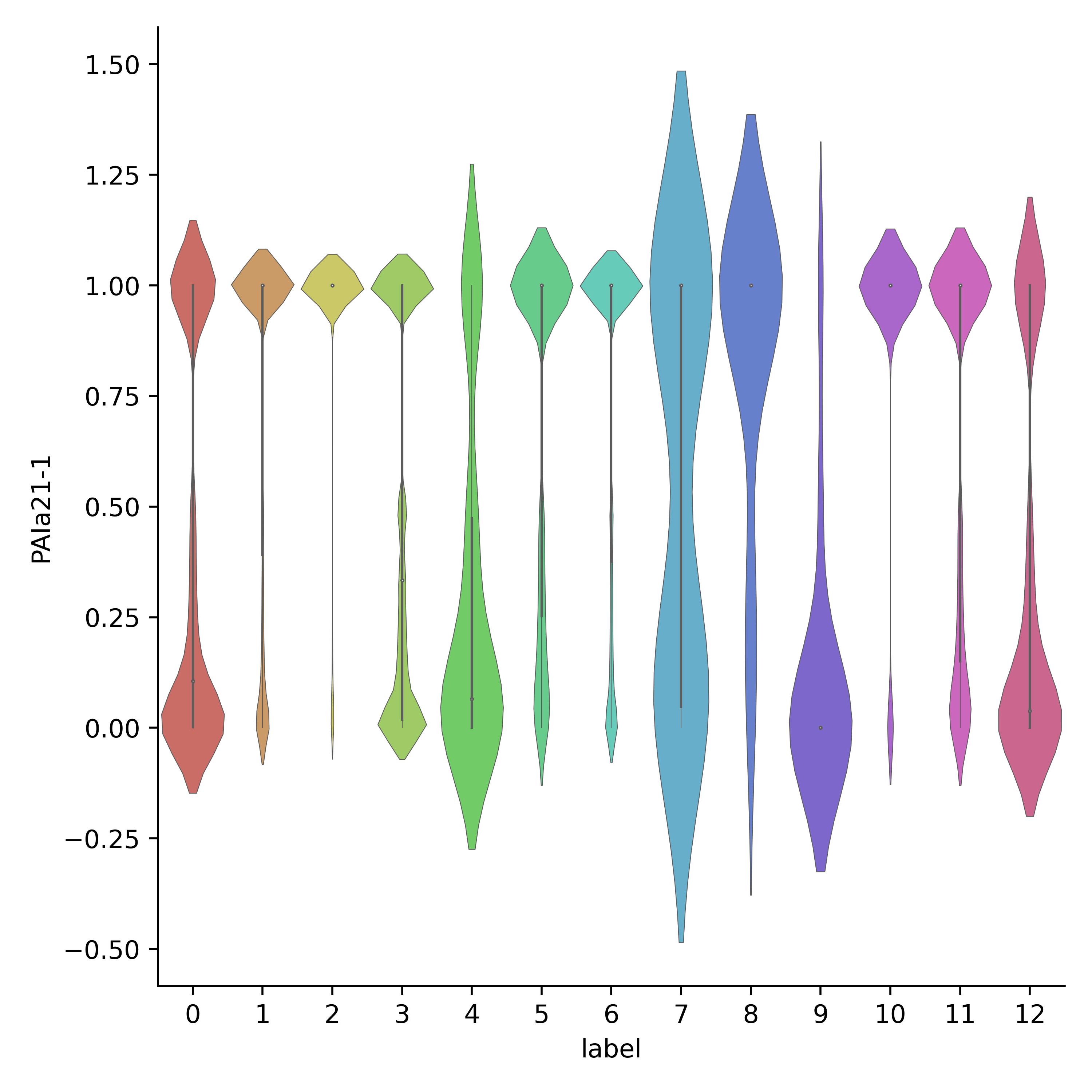}
\end{minipage}%
}%
\subfigure[PDIa1-2]{
\begin{minipage}[t]{0.3\linewidth}
\label{s2}
\centering
\includegraphics[width=1.75in]{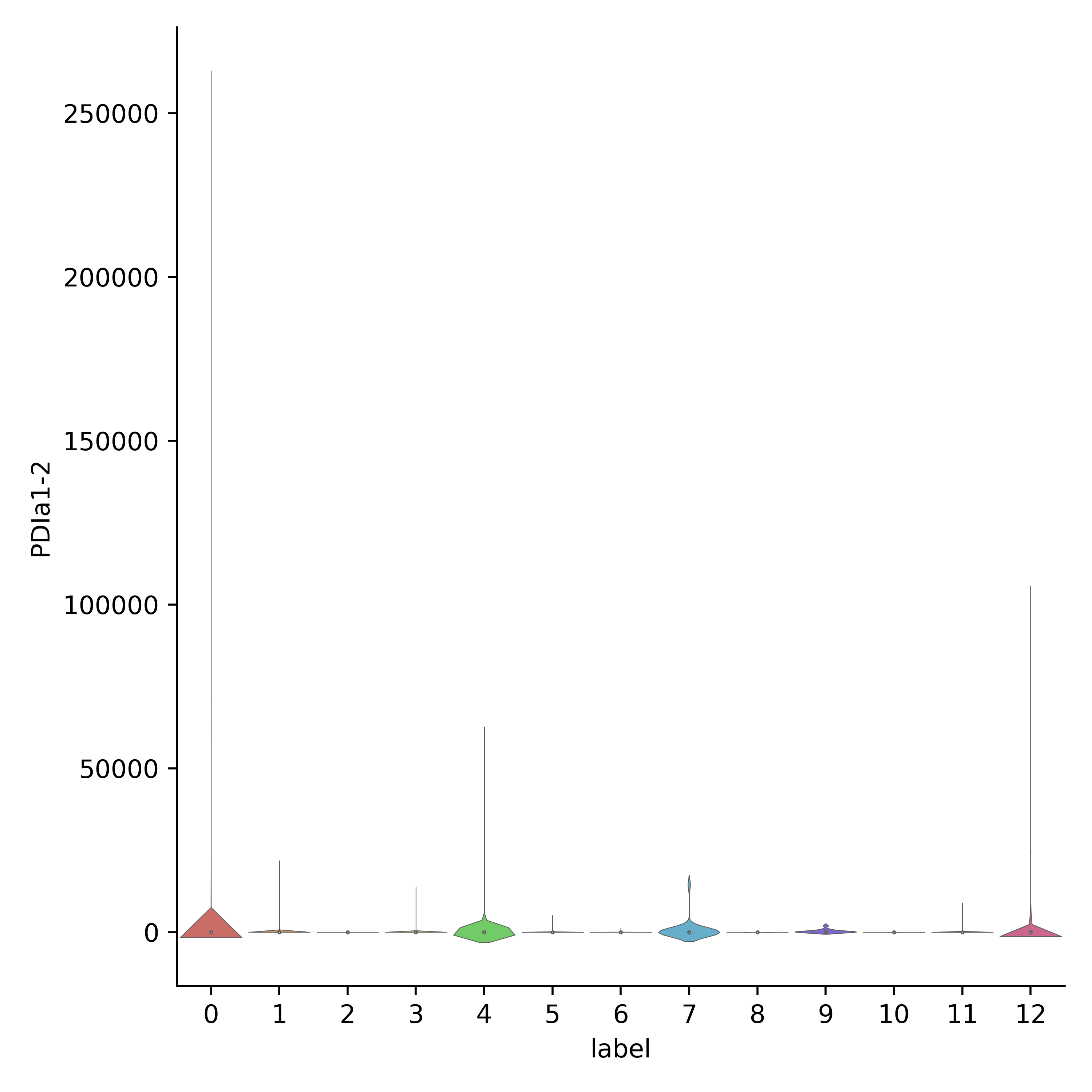}
\end{minipage}%
}%
\quad                 
\subfigure[PTIa41-2]{
\begin{minipage}[t]{0.3\linewidth}
\label{s3}
\centering
\includegraphics[width=1.75in]{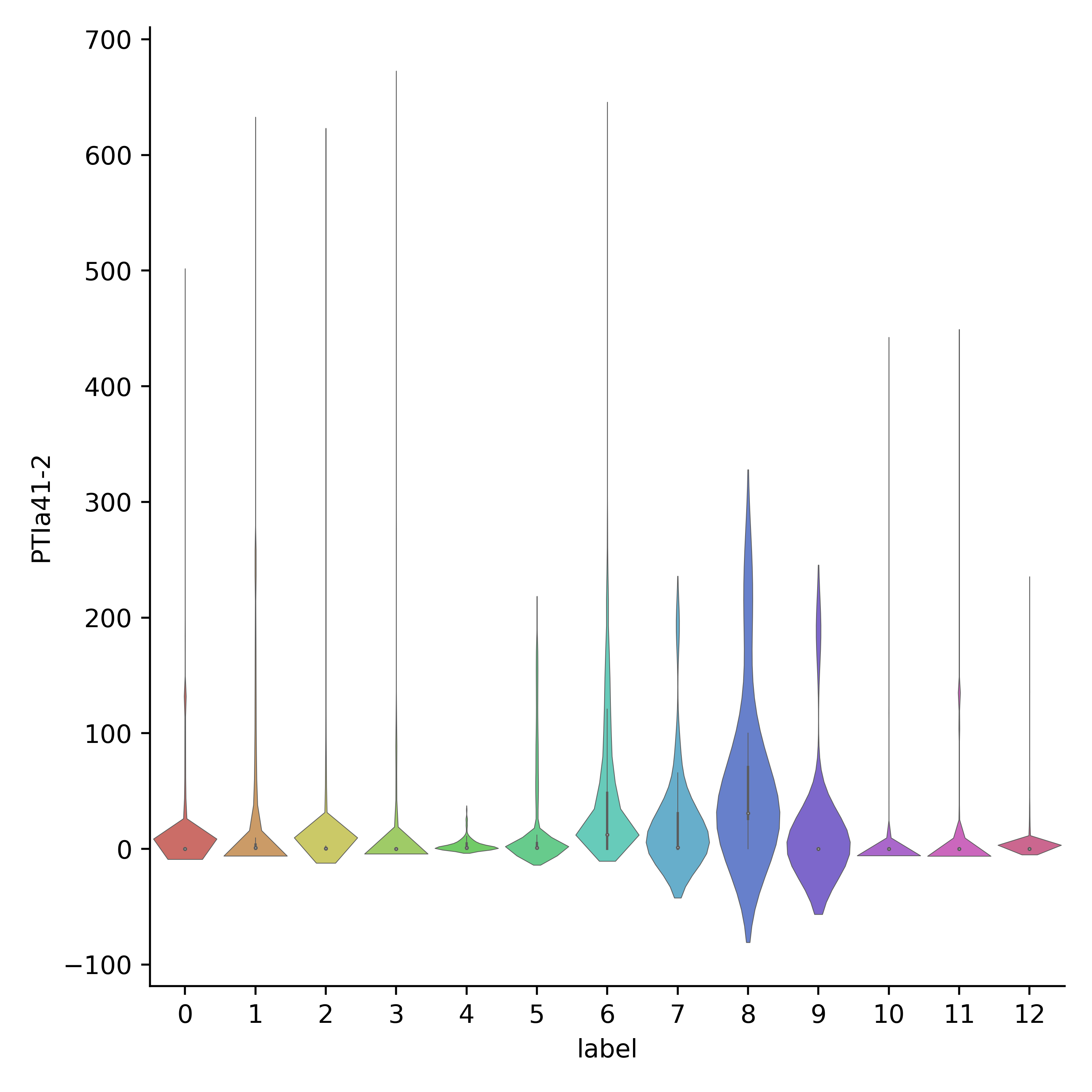}
\end{minipage}
} \quad%
\subfigure[CI3a32-2]{
\begin{minipage}[t]{0.3\linewidth}
\label{s4}
\centering
\includegraphics[width=1.75in]{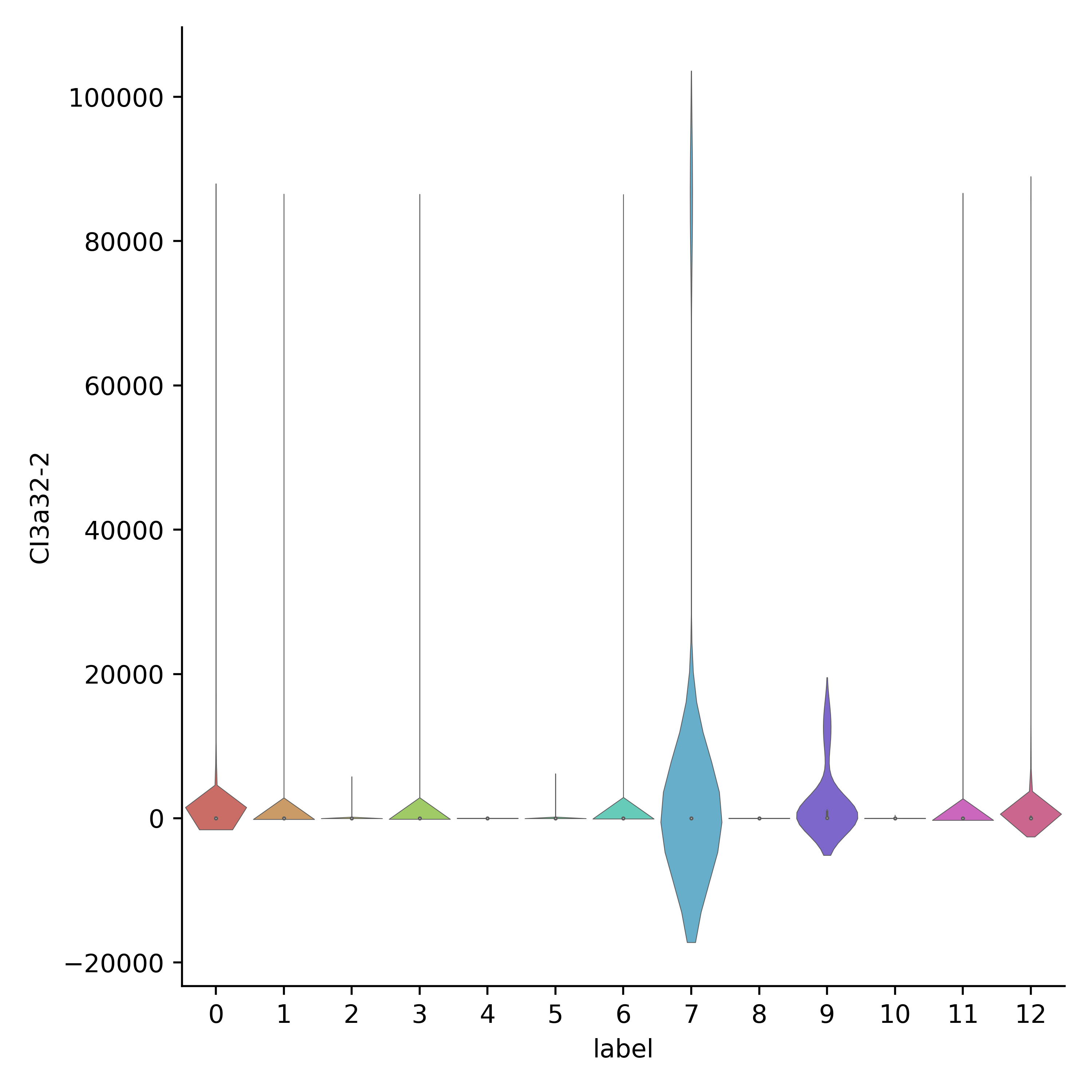}
\end{minipage}
}%
\subfigure[S2-2]{
\begin{minipage}[t]{0.3\linewidth}
\label{s5}
\centering
\includegraphics[width=1.75in]{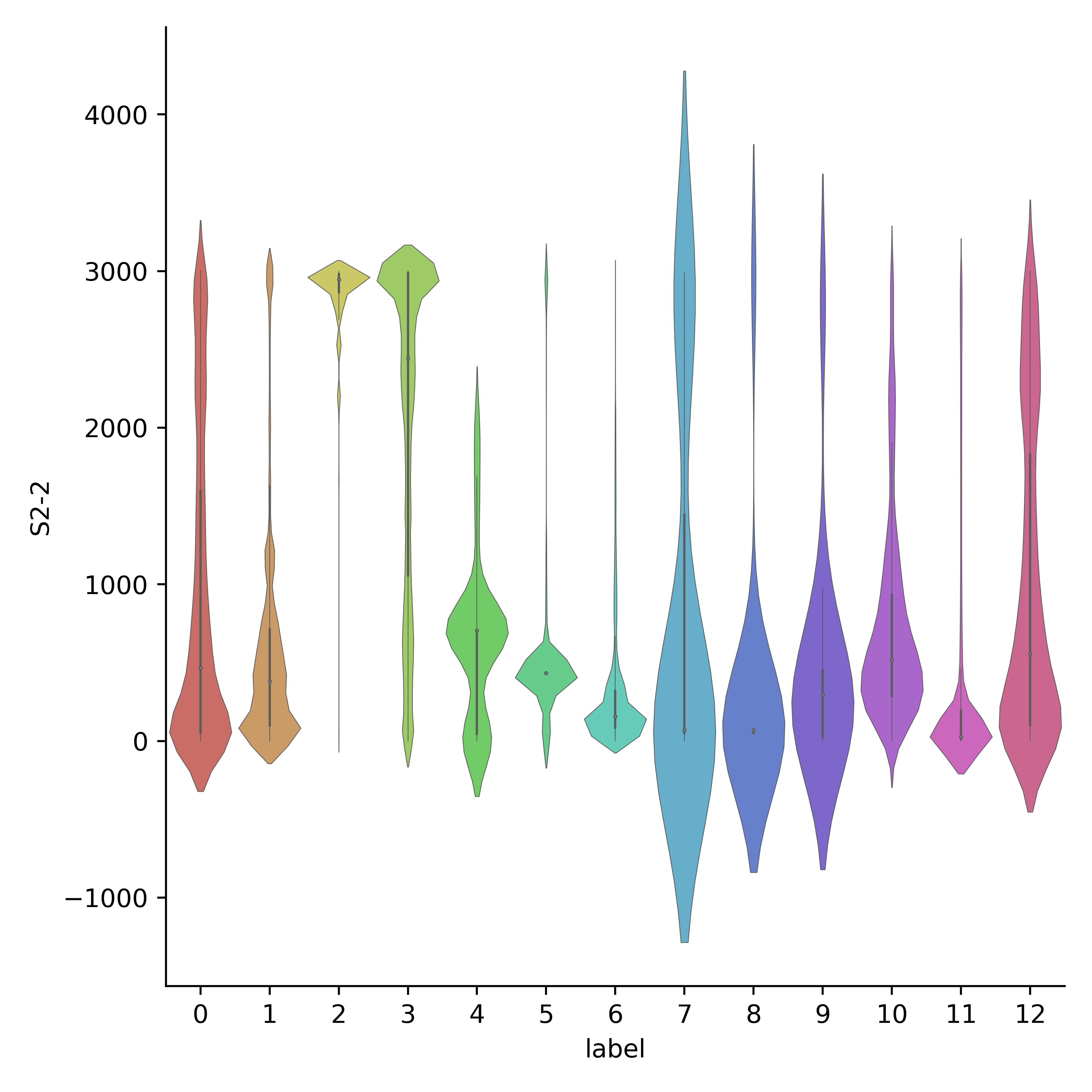}
\end{minipage}
}%
\subfigure[S2-3]{
\begin{minipage}[t]{0.3\linewidth}
\label{s6}
\centering
\includegraphics[width=1.75in]{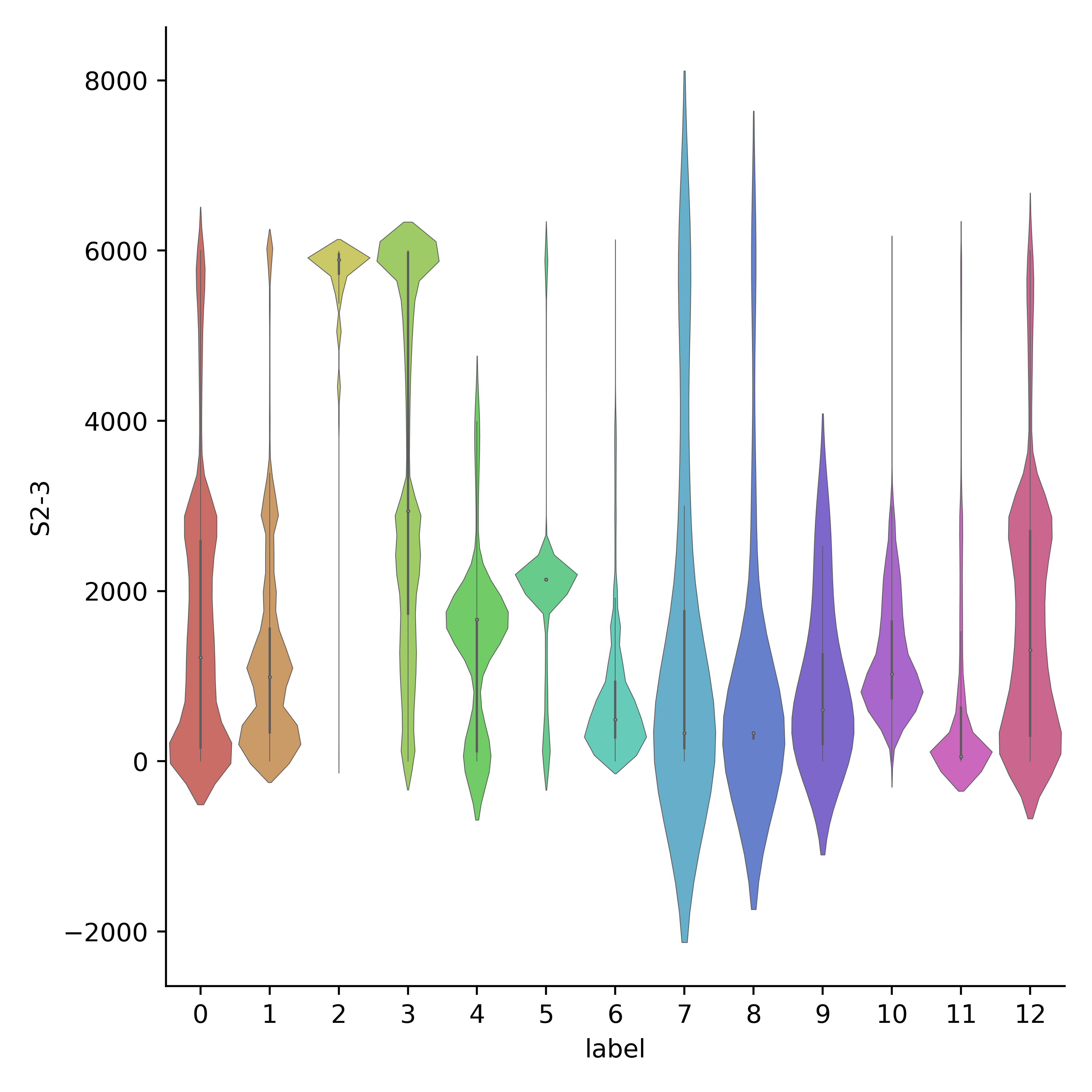}
\end{minipage}
}%
\centering
\label{v}
\caption{The violin plots of BABD-13 feature value distribution}
\end{figure*}

In Fig. \ref{s1}, the data distributions of all types of addresses have two peaks, furthermore, it is clear that the data distributions of all types of addresses on this feature are symmetric with different shapes and the symmetry axes are around 0.5.

For Fig. \ref{s2}, the data distributions are similar and have one peak that is not apparent. In addition, we can see that the data distributions of types 2, 5, 6, 8, and 9 are almost the same. We think that is the reason why this feature ranks low in the feature importance compared with other features selected in this part.

In Fig. \ref{s3}, we can observe that there are several types that are similar. For example, types 1, 3, 10, and 11. But other types have different distributions, where the data distribution of type 8 is the most distinct.

For Fig. \ref{s4}, the data distribution of types 7 and 8 are unique to other types. Besides, types 0 and 1 are similar; types 1, 3, 6, and 11are similar; types 2, 4, 5, 8, and 10 are similar. That might be the reason why this feature ranked not high in the top 10 influential features.

In Fig. \ref{s5} data distributions of all types have obvious different shapes which means it is easy to distinguish different types of addresses using this feature. Concretely, the differences show in many aspects, such as the number of peaks, the location of the peak, the shape of the distribution, and the symmetry.

Finally, we select a high correlation feature related to S2-2, i.e., S2-3, to compare their similarities and differences. In Fig. \ref{s5} and Fig. \ref{s6}, it is clear that most features with the same type have similar shapes. There are some with partially similar shapes, such as types 0, 1, and 12. That proves the reason why the correlation is high in S2-2 and S2-3 in the heatmap.

Moreover, based on the results above, we can raise some new valuable questions. For example, how to use the feature value distributions to extract novel meaningful features? And is it possible for an address to have more than one label due to there being overlapping in specific feature value distributions on different types of addresses? These issues we will continue exploring through BABD-13 in the near future.

\section{Discussion}
\subsection{Preliminary Analysis}
We use \emph{PDI} to show how we apply these features and the definitions of different Bitcoin address types to analyze behavior patterns. We choose darknet market, money laundering, and Ponzi with their total degrees (i.e. PDIa1-3) as examples.

We compute the averages, maximums, and medians of PDIa1-3 of these three types of addresses that are shown in Table \ref{analysis}. The median of PDIa1-3 of Ponzi is higher than others because Ponzi needs a lot of people to continually pay to an address to continue this fraud, while the frequency of money laundering is quite low for a single address because the person who wants to launder money needs to hide the movements of the Bitcoin flows. Besides, the transactions that happen in the darknet market are fewer due to the high threshold, high risk, and high cost.

\begin{table}[!htbp]
    \centering
    \caption{\emph{PDIa1-3} Analysis}
    \label{analysis}
    \begin{tabular}{l|rrrr}
        \toprule
        \diagbox{Type}{Value} &  AVG & MAX & MIN & MED\\
        \hline
        \midrule
        Darknet Market & 2.1 & 14 & 1 & 2 \\
        Money Laundering & 2 & 4 & 1 & 2 \\
        Ponzi & 606.4 & 3,939 & 2 & 74 \\
        \bottomrule
    \end{tabular}
\end{table}

\subsection{Dataset Improvement}
We can see from \ref{s2} that the effect of \emph{PDI} related features is bad in the classification task. In addition, according to Table \ref{analysis}, it is obvious that the information from \emph{PDI} related features we can obtain is limited. The reason is that the final status of \emph{PDI} related features is not the most important feature of Bitcoin address transaction. Also, \emph{PDI} related features are only the first layer of Bitcoin addresses that do not include the other layers.

In order to solve the above problem, we consider that it is possible to analyze the regular pattern of the growth of Bitcoin addresses in the same time interval and extract information from growth patterns as the new features that can be added to \emph{PDI}. The key issue of this study is how to choose the time interval and how to construct suitable subgraphs.

\section{Conclusion and Future Work}
In this paper, we proposed a framework for Bitcoin address behavior pattern analysis on the Bitcoin transaction graph constructed by the directed heterogeneous multigraph. The latter is built based on our Bitcoin address behavior dataset (BABD-13). We also proposed a novel approach to generate \emph{k}-hop subgraph for an address node from the entire Bitcoin transaction graph built by the directed heterogeneous multigraph in order to extract \emph{LSI}. We then used BABD-13 to evaluate 5 common machine learning models and achieved an accuracy rate of between 93.24\% and 97.13\%. If we wish to achieve better performance, we can also simplify BABD-13.

No work is perfect, and there are a number of potential research extensions. First, BABD-13 can be studied further in order to help us determine other useful features and characteristics and deeply mine the patterns of different types of Bitcoin addresses. Second, we are also exploring how we can extend our proposed framework from Bitcoin address behavioral analysis to Bitcoin transaction mode analysis, and use the combined framework to obtain a more comprehensive pattern framework or model. Finally, we also consider designing a method to identify entities, such as in the approaches outlined in \cite{Monaco,Chang,Zheng2,Yousaf,Nick,Shah,Shen}, from the directed heterogeneous multigraph Bitcoin transaction graph to distinguish and analyze different entities.

\section*{Acknowledgment}
The research was financially supported by the CCF-NSFOCUS Kun-Peng Scientific Research Fund (No. CCF-NSFOCUS2021008), the Provincial Key Research and Development Program of Hubei (No. 2020BAB105), the National Natural Science Foundation of China (No. 61972366), the Foundation of Henan Key Laboratory of Network Cryptography Technology (No. LNCT2020-A01), and the Foundation of Hubei Key Laboratory of Intelligent Geo-Information Processing (No. KLIGIP-2021B06). 

Also, we would like to thank Prof. Linchuan Xiang from the School of Physics, Huazhong University of Science and Technology, and Aleš Janda the founder of WalletExplorer. 

\bibliographystyle{IEEEtran}
\bibliography{sample}

\begin{thebibliography}{10}
\providecommand{\url}[1]{#1}
\csname url@samestyle\endcsname
\providecommand{\newblock}{\relax}
\providecommand{\bibinfo}[2]{#2}
\providecommand{\BIBentrySTDinterwordspacing}{\spaceskip=0pt\relax}
\providecommand{\BIBentryALTinterwordstretchfactor}{4}
\providecommand{\BIBentryALTinterwordspacing}{\spaceskip=\fontdimen2\font plus
\BIBentryALTinterwordstretchfactor\fontdimen3\font minus
  \fontdimen4\font\relax}
\providecommand{\BIBforeignlanguage}[2]{{%
\expandafter\ifx\csname l@#1\endcsname\relax
\typeout{** WARNING: IEEEtran.bst: No hyphenation pattern has been}%
\typeout{** loaded for the language `#1'. Using the pattern for}%
\typeout{** the default language instead.}%
\else
\language=\csname l@#1\endcsname
\fi
#2}}
\providecommand{\BIBdecl}{\relax}
\BIBdecl

\bibitem{Alqassem}
I.~Alqassem, I.~Rahwan, and D.~Svetinovic, ``The anti-social system properties:
  Bitcoin network data analysis,'' \emph{IEEE Transactions on Systems, Man, and
  Cybernetics: Systems}, vol.~50, no.~1, pp. 21--31, 2018.

\bibitem{Nerurkar}
P.~Nerurkar, D.~Patel, Y.~Busnel, R.~Ludinard, S.~Kumari, and M.~K. Khan,
  ``Dissecting bitcoin blockchain: Empirical analysis of bitcoin network
  (2009–2020),'' \emph{Journal of Network and Computer Applications}, vol.
  177, p. 102940, 2021.

\bibitem{Popuri}
M.~K. Popuri and M.~H. Gunes, \emph{Empirical analysis of crypto
  currencies}.\hskip 1em plus 0.5em minus 0.4em\relax Springer, 2016, pp.
  281--292.

\bibitem{Ron}
D.~Ron and A.~Shamir, ``Quantitative analysis of the full bitcoin transaction
  graph,'' in \emph{International Conference on Financial Cryptography and Data
  Security}.\hskip 1em plus 0.5em minus 0.4em\relax Springer, 2013, Conference
  Proceedings, pp. 6--24.

\bibitem{Serena}
L.~Serena, S.~Ferretti, and G.~D’Angelo, ``Cryptocurrencies activity as a
  complex network: Analysis of transactions graphs,'' \emph{Peer-to-Peer
  Networking and Applications}, pp. 1--15, 2021.

\bibitem{Tao}
B.~Tao, I.~W.-H. Ho, and H.-N. Dai, ``Complex network analysis of the bitcoin
  blockchain network,'' in \emph{2021 IEEE International Symposium on Circuits
  and Systems (ISCAS)}.\hskip 1em plus 0.5em minus 0.4em\relax IEEE, Conference
  Proceedings, pp. 1--5.

\bibitem{tao2021complex}
B.~Tao, H.-N. Dai, J.~Wu, I.~W.-H. Ho, Z.~Zheng, and C.~F. Cheang, ``Complex
  network analysis of the bitcoin transaction network,'' \emph{IEEE
  Transactions on Circuits and Systems II: Express Briefs}, 2022.

\bibitem{Ranshous}
S.~Ranshous, C.~A. Joslyn, S.~Kreyling, K.~Nowak, N.~F. Samatova, C.~L. West,
  and S.~Winters, ``Exchange pattern mining in the bitcoin transaction directed
  hypergraph,'' in \emph{International Conference on Financial Cryptography and
  Data Security}.\hskip 1em plus 0.5em minus 0.4em\relax Springer, Conference
  Proceedings, pp. 248--263.

\bibitem{Romiti}
M.~Romiti, A.~Judmayer, A.~Zamyatin, and B.~Haslhofer, ``A deep dive into
  bitcoin mining pools: An empirical analysis of mining shares,'' \emph{arXiv
  preprint arXiv:1905.05999}, 2019.

\bibitem{Tovanich}
N.~Tovanich, N.~Soulié, N.~Heulot, and P.~Isenberg, ``An empirical analysis of
  pool hopping behavior in the bitcoin blockchain,'' in \emph{2021 IEEE
  International Conference on Blockchain and Cryptocurrency}, Conference
  Proceedings.

\bibitem{Li}
Y.~Li, Y.~Cai, H.~Tian, G.~Xue, and Z.~Zheng, ``Identifying illicit addresses
  in bitcoin network,'' in \emph{International Conference on Blockchain and
  Trustworthy Systems}.\hskip 1em plus 0.5em minus 0.4em\relax Springer,
  Conference Proceedings, pp. 99--111.

\bibitem{Paquet-Clouston1}
M.~Paquet-Clouston, M.~Romiti, B.~Haslhofer, and T.~Charvat, ``Spams meet
  cryptocurrencies: Sextortion in the bitcoin ecosystem,'' in \emph{Proceedings
  of the 1st ACM conference on advances in financial technologies}, Conference
  Proceedings, pp. 76--88.

\bibitem{Weber}
M.~Weber, G.~Domeniconi, J.~Chen, D.~K.~I. Weidele, C.~Bellei, T.~Robinson, and
  C.~E. Leiserson, ``Anti-money laundering in bitcoin: Experimenting with graph
  convolutional networks for financial forensics,'' \emph{arXiv preprint
  arXiv:1908.02591}, 2019.

\bibitem{Wu3}
J.~Wu, J.~Liu, W.~Chen, H.~Huang, Z.~Zheng, and Y.~Zhang, ``Detecting mixing
  services via mining bitcoin transaction network with hybrid motifs,''
  \emph{IEEE Transactions on Systems, Man, and Cybernetics: Systems}, 2021.

\bibitem{Liao}
K.~Liao, Z.~Zhao, A.~Doupé, and G.-J. Ahn, ``Behind closed doors: measurement
  and analysis of cryptolocker ransoms in bitcoin,'' in \emph{2016 APWG
  symposium on electronic crime research (eCrime)}.\hskip 1em plus 0.5em minus
  0.4em\relax IEEE, Conference Proceedings, pp. 1--13.

\bibitem{Conti}
M.~Conti, A.~Gangwal, and S.~Ruj, ``On the economic significance of ransomware
  campaigns: A bitcoin transactions perspective,'' \emph{Computers \&
  Security}, vol.~79, pp. 162--189, 2018.

\bibitem{Paquet-Clouston}
M.~Paquet-Clouston, B.~Haslhofer, and B.~Dupont, ``Ransomware payments in the
  bitcoin ecosystem,'' \emph{Journal of Cybersecurity}, vol.~5, no.~1, p.
  tyz003, 2019.

\bibitem{Bartoletti}
M.~Bartoletti, B.~Pes, and S.~Serusi, ``Data mining for detecting bitcoin ponzi
  schemes,'' in \emph{2018 Crypto Valley Conference on Blockchain Technology
  (CVCBT)}.\hskip 1em plus 0.5em minus 0.4em\relax IEEE, 2018, Conference
  Proceedings, pp. 75--84.

\bibitem{Vasek}
M.~Vasek and T.~Moore, ``Analyzing the bitcoin ponzi scheme ecosystem,'' in
  \emph{International Conference on Financial Cryptography and Data
  Security}.\hskip 1em plus 0.5em minus 0.4em\relax Springer, Conference
  Proceedings, pp. 101--112.

\bibitem{Toyoda}
K.~Toyoda, P.~T. Mathiopoulos, and T.~Ohtsuki, ``A novel methodology for hyip
  operators’ bitcoin addresses identification,'' \emph{IEEE Access}, vol.~7,
  pp. 74\,835--74\,848, 2019.

\bibitem{Liu1}
X.~F. Liu, X.-J. Jiang, S.-H. Liu, and C.~K. Tse, ``Knowledge discovery in
  cryptocurrency transactions: A survey,'' \emph{IEEE Access}, vol.~9, pp.
  37\,229--37\,254, 2021.

\bibitem{Monaco}
J.~V. Monaco, ``Identifying bitcoin users by transaction behavior,'' in
  \emph{Biometric and Surveillance Technology for Human and Activity
  Identification XII}, vol. 9457.\hskip 1em plus 0.5em minus 0.4em\relax
  International Society for Optics and Photonics, 2015, Conference Proceedings,
  p. 945704.

\bibitem{Chang}
T.-H. Chang and D.~Svetinovic, ``Improving bitcoin ownership identification
  using transaction patterns analysis,'' \emph{IEEE Transactions on Systems,
  Man, and Cybernetics: Systems}, vol.~50, no.~1, pp. 9--20, 2018.

\bibitem{Zola}
F.~Zola, J.~L. Bruse, M.~Eguimendia, M.~Galar, and R.~Orduna~Urrutia, ``Bitcoin
  and cybersecurity: temporal dissection of blockchain data to unveil changes
  in entity behavioral patterns,'' \emph{Applied Sciences}, vol.~9, no.~23, p.
  5003, 2019.

\bibitem{8751410}
Y.-J. Lin, P.-W. Wu, C.-H. Hsu, I.-P. Tu, and S.-w. Liao, ``An evaluation of
  bitcoin address classification based on transaction history summarization,''
  in \emph{2019 IEEE International Conference on Blockchain and Cryptocurrency
  (ICBC)}, 2019, pp. 302--310.

\bibitem{Wu2}
J.~Wu, J.~Liu, Y.~Zhao, and Z.~Zheng, ``Analysis of cryptocurrency transactions
  from a network perspective: An overview,'' \emph{Journal of Network and
  Computer Applications}, p. 103139, 2021.

\bibitem{Zheng2}
B.~Zheng, L.~Zhu, M.~Shen, X.~Du, and M.~Guizani, ``Identifying the
  vulnerabilities of bitcoin anonymous mechanism based on address clustering,''
  \emph{Science China Information Sciences}, vol.~63, no.~3, pp. 1--15, 2020.

\bibitem{Yousaf}
H.~Yousaf, G.~Kappos, and S.~Meiklejohn, ``Tracing transactions across
  cryptocurrency ledgers,'' in \emph{28th {USENIX} Security Symposium ({USENIX}
  Security 19)}, 2019, Conference Proceedings, pp. 837--850.

\bibitem{Maesa}
D.~Di~Francesco~Maesa, A.~Marino, and L.~Ricci, ``Uncovering the bitcoin
  blockchain: An analysis of the full users graph,'' in \emph{2016 IEEE
  International Conference on Data Science and Advanced Analytics (DSAA)},
  2016, pp. 537--546.

\bibitem{Liu}
X.~F. Liu, H.-H. Ren, S.-H. Liu, and X.-J. Jiang, ``Characterizing key agents
  in the cryptocurrency economy through blockchain transaction analysis,''
  \emph{EPJ Data Science}, vol.~10, no.~1, pp. 1--13, 2021.

\bibitem{Greaves}
A.~Greaves and B.~Au, ``Using the bitcoin transaction graph to predict the
  price of bitcoin,'' \emph{No Data}, 2015.

\bibitem{Boccaletti1}
S.~Boccaletti, V.~Latora, Y.~Moreno, M.~Chavez, and D.-U. Hwang, ``Complex
  networks: Structure and dynamics,'' \emph{Physics Reports}, vol. 424, no.~4,
  pp. 175--308, 2006.

\bibitem{Newman}
M.~E.~J. Newman, ``Assortative mixing in networks,'' \emph{Phys. Rev. Lett.},
  vol.~89, p. 208701, Oct 2002.

\bibitem{okamoto}
K.~Okamoto, W.~Chen, and X.-Y. Li, ``Ranking of closeness centrality for
  large-scale social networks,'' in \emph{International workshop on frontiers
  in algorithmics}.\hskip 1em plus 0.5em minus 0.4em\relax Springer, 2008, pp.
  186--195.

\bibitem{Lawrence}
L.~Page, S.~Brin, R.~Motwani, and T.~Winograd, ``The pagerank citation ranking
  : Bringing order to the web,'' in \emph{WWW 1999}, 1999.

\bibitem{bollobas1998modern}
B.~Bollob{\'a}s and B.~Bollobas, \emph{Modern graph theory}.\hskip 1em plus
  0.5em minus 0.4em\relax Springer Science \& Business Media, 1998, vol. 184.

\bibitem{peixoto_graph-tool_2014}
\BIBentryALTinterwordspacing
T.~P. Peixoto, ``The graph-tool python library,'' \emph{figshare}, 2014.
  [Online]. Available: \url{http://figshare.com/articles/graph_tool/1164194}
\BIBentrySTDinterwordspacing

\bibitem{Hagberg}
A.~Hagberg, P.~Swart, and D.~S~Chult, ``Exploring network structure, dynamics,
  and function using networkx,'' Los Alamos National Lab.(LANL), Los Alamos, NM
  (United States), Report, 2008.

\bibitem{scikit-learn}
F.~Pedregosa, G.~Varoquaux, A.~Gramfort, V.~Michel, B.~Thirion, O.~Grisel,
  M.~Blondel, P.~Prettenhofer, R.~Weiss, V.~Dubourg, J.~Vanderplas, A.~Passos,
  D.~Cournapeau, M.~Brucher, M.~Perrot, and E.~Duchesnay, ``Scikit-learn:
  Machine learning in {P}ython,'' \emph{Journal of Machine Learning Research},
  vol.~12, pp. 2825--2830, 2011.

\bibitem{xgboost}
T.~Chen and C.~Guestrin, ``{XGBoost}: A scalable tree boosting system,'' in
  \emph{Proceedings of the 22nd ACM SIGKDD International Conference on
  Knowledge Discovery and Data Mining}, ser. KDD '16.\hskip 1em plus 0.5em
  minus 0.4em\relax New York, NY, USA: ACM, 2016, pp. 785--794.

\bibitem{michalski2020revealing}
R.~Michalski, D.~Dziubałtowska, and P.~Macek, ``Revealing the character of
  nodes in a blockchain with supervised learning,'' \emph{IEEE Access}, vol.~8,
  pp. 109\,639--109\,647, 2020.

\bibitem{Nick}
J.~D. Nick, ``Data-driven de-anonymization in bitcoin,'' Thesis, 2015.

\bibitem{Shah}
R.~S. Shah, A.~Bhatia, A.~Gandhi, and S.~Mathur, ``Bitcoin data analytics:
  Scalable techniques for transaction clustering and embedding generation,'' in
  \emph{2021 International Conference on COMmunication Systems \& NETworkS
  (COMSNETS)}.\hskip 1em plus 0.5em minus 0.4em\relax IEEE, 2021, Conference
  Proceedings, pp. 1--6.

\bibitem{Shen}
M.~Shen, J.~Duan, N.~Shang, and L.~Zhu, ``Transaction deanonymization in
  large-scale bitcoin systems via propagation pattern analysis,'' in
  \emph{International Conference on Security and Privacy in Digital
  Economy}.\hskip 1em plus 0.5em minus 0.4em\relax Springer, 2020, Conference
  Proceedings, pp. 661--675.

\end{thebibliography}

\end{document}